\RequirePackage{ifpdf}
\documentclass[12pt,letterpaper]{article}
\pdfoutput=1
\usepackage{jheppub}
\usepackage{epsfig}
\usepackage{enumitem}
\usepackage[utf8x]{inputenc}
\usepackage{bbm,amsfonts}
\usepackage{graphicx}
\usepackage{amssymb,amsmath,amsfonts,mathtools,psfrag}
\usepackage{fancybox}
\usepackage{enumerate}
\usepackage{dsfont}
\usepackage{verbatim}

\usepackage{etex}
\usepackage{pstricks-add}
\usepackage{verbatim}

\usepackage{bm}
\usepackage{epic}
\usepackage{color}
\usepackage{young}
\usepackage{multirow}
\usepackage{subfig}
\usepackage{float}

\def \be  {\begin{equation}}
\def \ee  {\end{equation}}
\def \ba  {\begin{eqnarray}}
\def \ea  {\end{eqnarray}}
\def \baa {\begin{eqnarray*}}
\def \eaa {\end{eqnarray*}}
\def \bb  {\begin {thebibliography} }
\def \eb  {\end{thebibliography}}
\def \lab #1 {\label{#1}}

\newcommand {\non}{\nonumber}

\def\tr{\mathrm{Tr}}
\def\d{\mathrm{d}}

\def\th{\theta}

\def\bseq{\begin{subequation}}  
\def\eseq{\end{subequation}}
\def\bsea{\begin{subeqnarray}}  
\def\esea{\end{subeqnarray}}

\newcommand{\beq}{\begin{equation}}
\newcommand{\bea}{\begin{eqnarray}}
\newcommand{\eea}{\end{eqnarray}}
\newcommand{\eeq}{\end{equation}}

\renewcommand{\a}{\alpha}
\renewcommand{\b}{\beta}

\newcommand{\e}{\epsilon}
\newcommand{\z}{\zeta}

\renewcommand{\l}{\lambda}

\newcommand{\m}{\mu}

\newcommand{\n}{\nu}

\renewcommand{\t}{\tau}

\begin{document}

\title{The Bremsstrahlung function of $\mathcal{N} \!= \!2 $  SCQCD}

\author[a,b]{Carolina Gomez,}
\author[a]{Andrea Mauri,}
\author[a,b]{ Silvia Penati}

\affiliation[a]{ Dipartimento di Fisica, Universit\`a degli studi di Milano--Bicocca, Piazza della Scienza 3, I-20126 Milano, Italy }
\affiliation[b]{INFN, Sezione di Milano--Bicocca, Piazza della Scienza 3, I-20126 Milano, Italy }

\abstract{For $SU(N)$ superconformal QCD we perform a three--loop calculation of the cusp anomalous dimension for a generalized Maldacena--Wilson operator, using HQET formalism.  We obtain an expression that is valid at generic geometric and internal angles and finite gauge group rank $N$. For equal and opposite angles this expression vanishes, proving that at these points the cusp becomes BPS. From its small angle expansion we derive the corresponding Bremsstrahlung function at three loops, matching the matrix model prediction given in terms of derivatives of the Wilson loop on the ellipsoid. Finally, we discuss  possible scenarios at higher loops,  with respect to the existence of a universal effective coupling in an integrable subsector of the model.}

\emailAdd{carolina.gomez@mib.infn.it} 
\emailAdd{andrea.mauri@mi.infn.it} 
\emailAdd{silvia.penati@mib.infn.it}


\maketitle

\section{Introduction} \label{section1}

The Bremsstrahlung function $B$ is a physical quantity that plays an ubiquitous role when probing quantum field theories. It is defined as the energy lost by a heavy quark slowly moving in a gauge background  \cite{Correa:2012at}
\be
\Delta E = 2\pi B \, \int dt (\dot{v})^2, \qquad |v| \ll 1
\ee
and generalizes the well known constant $B = e^2/3\pi$ of electrodynamics.
 
In a conformal field theory, it also coincides with the lowest order coefficient in the small angle expansion of the cusp anomalous dimension 
\begin{equation}\label{B}
 \Gamma(\varphi, g) \underset{\varphi  \ll 1}{\sim} -B(g)\varphi^2 
\end{equation}
which governs the short distance behavior of a $\varphi$--cusped Wilson operator  
\begin{equation} \label{CAD}
 \langle W\rangle\sim e^{-\Gamma(\varphi, g) \log \frac{\Lambda}{\mu}}
\end{equation}
Here $\Lambda$ and $\mu$ are the IR cut--off and the UV renormalization scale, respectively, whereas $\Gamma$ is a function of the geometric angle of the cusp and the coupling $g$ of the theory. Consequently, $B$ in \eqref{B} is in general a non--trivial function of $g$. 

Equation \eqref{CAD} provides the standard prescription for computing the Bremsstrahlung function at weak coupling. In fact, it is sufficient to compute the cusped Wilson operator order by order in $g$, using dimensional or cut--off regularization to tame UV divergences and introducing a suppression factor to mitigate divergences at large distances. After removing the IR regulator by a multiplicative renormalization ($\langle W\rangle \to \langle \widetilde W\rangle$, see eq. \eqref{IRremoving}) and renormalizing the short distance divergences at scale $\mu$
\begin{equation}
 \langle W_R\rangle=Z^{-1}_{cusp} \langle\widetilde{W}\rangle \qquad\mathrm{s.t.}\qquad \frac{d\log\langle W_R\rangle}{d\log\mu}=0
\end{equation}
we finally read $\Gamma$ as
\begin{equation}
 \Gamma(\varphi, g) =\frac{d\log(Z_{cusp})}{d\log\mu}\;,
 \label{gam}
\end{equation}
Its small angle expansion then leads to $B$ at a given order in the coupling, according to eq. \eqref{B}. In dimensional regularization, which will be used in this paper, $\Gamma$ can be read from the coefficient of the $1/\epsilon$ pole.  

In order to use $B$ for probing the theory at different scales, aimed for instance at performing precision tests of the AdS/CFT correspondence, one needs to go beyond the perturbative regime. A clever way to do that is to relate $B$ to quantities that can be computed holographically and, in superconformal theories, by the use of localization techniques. The top candidates for these quantities are circular BPS Wilson loops for which exact results can be obtained from a computable matrix model. 

For ${\cal N} = 4$ $SU(N)$ SYM theory, in \cite{Correa:2012at} it was proved that the Bremsstrahlung function can be computed as a derivative of the vacuum expectation value (vev) of a circular 1/2 BPS Wilson loop with respect to the 't Hooft coupling $\lambda \equiv g^2 N$ 
\begin{equation}
 B_{\mathcal{N}=4}=\frac{1}{2\pi^2}\lambda\partial_{\lambda}\log\langle W\rangle
 \label{B4lambda}
\end{equation}
where $\langle W\rangle$ is computed exactly by a gaussian matrix model that localizes the vev on the four sphere $S^4$ \cite{Erickson:2000af, Drukker:2000rr, Pestun:2007rz}. Alternatively, working on an ellipsoid $S_b$ with squashing parameter $b$, the prescription for obtaining the Bremsstrahlung function takes the form
\begin{equation}
 \left.B_{\mathcal{N}=4}=\frac{1}{4\pi^2}\partial_{b}\log \langle W_b\rangle\right\vert_{b=1}
 \label{B4b}
\end{equation}
where $\langle W_b\rangle$ is the circular Wilson loop computed by the matrix model on the ellipsoid \cite{Hama:2012bg},\cite{Alday:2009fs},\cite{Fucito:2015ofa}.

A similar prescription has been conjectured \cite{Bianchi:2014laa} and then proved \cite{Bianchi:2017svd, Bianchi:2017ozk, Bianchi:2017afp, Bianchi:2017ujp, Bianchi:2018scb} for the $B$ function in three dimensional Chern--Simons--matter theories, notably the ABJ(M) model, where $B$ is related to the derivative of a latitude Wilson loop on $S^3$ respect to the latitude parameter. A matrix model for computing this quantity has been recently proposed in \cite{Bianchi:2018bke}. 

In this letter we are interested in four dimensional ${\cal N} = 2$ SYM theories, in particular ${\cal N} = 2$ $SU(N)$ superconformal QCD (SCQCD), for which Fiol, Gerchkovitz and Komargodski \cite{Fiol:2015spa}, inspired by the ${\cal N} = 4$ result, conjectured that  
\begin{equation} \label{B2}
 \left.B_{\mathcal{N}=2}=\frac{1}{4\pi^2}\partial_b\log\langle W_b\rangle\right\vert_{b=1}
\end{equation}
where again $\langle W_b\rangle$ is the circular Wilson loop corresponding to the matrix model on the ellipsoid.

Identity \eqref{B2} has been explicitly checked up to three loop for gauge group $SU(2)$  \cite{Fiol:2015spa}, while for $N>2$ only a consistency check of its positivity has been given there.  
One of the main goals of this letter is to extend this proof to the general $SU(N)$ case\footnote{Conjecture \eqref{B2} came together with a related one stating that $B_{\mathcal{N}=2}= 3h$, where $h$ is the coefficient of the one--point correlation function for the stress--energy tensor in the presence of the Wilson line defect. This second conjecture was later proved in \cite{Bianchi:2018zpb}.}. 

\vskip 10pt
To this end we consider a generalized Maldacena--Wilson operator  \cite{Maldacena:1998im} along a cusped line with geometric angle 
 $\varphi$ and featured by an internal angle $\theta$ which rotates the couplings to the adjoint matter when moving through the cusp. The corresponding generalized cusp anomalous dimension turns out to be a function of both angles,  $\Gamma(\varphi, \theta, g)$.  

For generic $SU(N)$ SCQCD we perform a genuine three--loop calculation of the cusped operator at generic angles and finite group rank $N$. From the $1/\epsilon$ pole of the dimensionally regularized result we then extract the generalized cusp $\Gamma(\varphi , \theta,g)$ at three loops (order $g^6$) and the corresponding $B$ from its small angle expansion. We find a general result that, remarkably, coincides with the r.h.s. of eq. \eqref{B2} once we expand the matrix model defining $\langle W_b\rangle$ up to ${\cal O}(g^6)$. This confirms the validity of conjecture \eqref{B2} for any $SU(N)$ gauge group.

Beyond providing a three loop check of this conjecture, our results (\ref{Gfinal}, \ref{Bfinal}) represent the first complete $\mathcal{N} = 2$ SCQCD corrections to $\Gamma(\varphi,\theta,g)$ and $B(g)$ at three loops. In particular, up to three loops we find that for small angles $\Gamma(\varphi, \theta,g) \sim B(g) (\theta^2 - \varphi^2)$ and the cusp vanishes at $\theta=\pm \varphi $. We then conclude that at these points the cusped Wilson operator becomes BPS, in analogy with the corresponding operator in ${\cal N}=4$ SYM theory. In fact, it can be proved that for  
$\theta=\pm \varphi$ the operator reduces to a Zarembo's type one \cite{Zarembo:2002an}. The left and the right rays of the cusp share the same superconformal charges, which are then globally preserved.

The cusp anomalous dimension for ordinary Wilson operators with no coupling to matter has been already computed up to three loops for QCD and supersymmetric gauge theories with matter in the adjoint representation  \cite{Grozin:2014hna, Grozin:2015kna}\footnote{For a summary of some partial four--loop  results see \cite{Bruser:2018aud} and references therein.}. Our result completes this picture by including both fundamental and adjoint matter in a supersymmetric way and considering BPS Wilson operators with non--trivial matter couplings.

In approaching the problem we use a different computational setup from the one used in  \cite{Fiol:2015spa}, where the  expression of the three--loop Bremsstrahlung function was derived by inserting in the cusp the resummed two--loop propagators as computed in \cite{Andree:2010na}. Rather we treat each diagram contributing to the cusp, separately -- no resummation involved -- and compute each of them using the Heavy Quark Effective Theory (HQET) formalism. This approach has the great advantage that, by applying a clever chain of integration by parts,  all the integrals can be expressed in terms of a linear combination of a basis of known three-loop HQET Master Integrals. In addition, it provides a promising framework where we can attempt higher--loop calculations and speculate about the origin of some unexpected terms in the higher order expansion of the B function \cite{Mitev:2015oty}, which can be shown to arise naturally in the HQET context.  

This computational set--up could be fruitfully used also for performing higher--loop tests of correlation functions in ${\cal N} = 2$ SCQCD and in the defect field theory defined on the Wilson contour, along the lines of \cite{Baggio:2014ioa, Billo:2017glv, Billo:2018oog, Beccaria:2018xxl, Beccaria:2018owt}. 

The paper is organized as follows. We first fix our conventions and describe our computational strategy in Section \ref{section:difference}, and recall the Matrix Model result in Section \ref{MM}. Then the core of the paper follows, where we report the diagrammatic approach to the three--loop calculation in section \ref{section3} and the HQET evaluation of the Feynman integrals in section \ref{section4}. The main results are presented in section \ref{section5} where we discuss the consistency of our findings with the conjectural expression \eqref{B2} for $SU(N)$ ${\cal N}=2$ SCQCD at any finite $N$. More generally,  we give the first complete three--loop expression for the ${\cal N} = 2$ corrections to the generalized cusp anomalous dimension and the corresponding Bremsstrahlung function. We also provide an explicit check of the universal behavior of the cusp anomalous dimension proposed in \cite{Grozin:2014hna, Grozin:2015kna}, which should work up to three loops. Finally, a critical discussion about the use of our technologies for going to higher loops is presented in section \ref{section6}. Appendix \ref{appendix:conventions} fixes the conventions needed to follow our calculations and Appendix \ref{appendix:partial} collects several computational details.

\section{The difference method}
\label{section:difference}

We will compute the cusp anomalous dimension and the associated Bremsstrahlung function of $\mathcal{N}=2$ SCQCD by comparing them with the corresponding known quantities of $\mathcal{N}=4$ SYM. In fact, it is well--known that this trick drastically reduces the number of new diagrams to be computed, as we briefly review here.

As a starting point, we find convenient to approach the problem within the framework of $\mathcal{N}=1$ superspace (see appendix \ref{appendix:conventions} for conventions).  In this language the field content of the $\mathcal{N}=2$ SCQCD theory with gauge group $SU(N)$ is organized into one vector and one chiral multiplets transforming in the adjoint representation of $SU(N)$, which form the $\mathcal{N}=2$ vector multiplet, together with $N_f = 2N$ chiral multiplets $Q_I, \bar{\tilde Q}_I$, $I = 1, \dots , 2 N$,  building up $2N$ $\mathcal{N}=2$ hypermultiplets transforming in the fundamental representation of the gauge group. 

Analogously, the $\mathcal{N}=4$ SYM theory is described by one vector multiplet plus a $SU(3)$ triplet of adjoint chiral multiplets. Together they build up the $\mathcal{N}=2$ vector multiplet, combining the $\mathcal{N}=1$ vector with one of the chiral multiplets  in analogy with the $\mathcal{N}=2$ SCQCD case,  plus one adjoint  $\mathcal{N}=2$ hypermultiplet from the two  remaining adjoint chiral multiplets. 

Therefore, the two theories have the same $\mathcal{N}=2$ gauge sector, while the difference relies only in the matter content and entails the comparison of two of the adjoint chiral superfields in $\mathcal{N}=4$ SYM as opposed to the pair of $2N$ superquark fundamentals in $\mathcal{N}=2$ SCQCD \cite{Rey:2010ry}. 
This allows to drastically simplify the calculation of any observable $O$ that is common to $\mathcal{N}=2$ SCQCD and $\mathcal{N}=4$ SYM theories if, instead of computing $\langle O\rangle_{\mathcal{N}=2}$ directly, one computes the difference $\langle O\rangle_{\mathcal{N}=2} - \langle O\rangle_{\mathcal{N}=4}$. In fact, in the difference all the Feynman diagrams that are common to the two theories cancel, in particular the ones built with fields belonging to the gauge sector.  The computational strategy of taking the difference was first introduced in \cite{Andree:2010na}, albeit working with a different description of the field content of the two theories.

We will work in the component formulation of the two models directly derived from projecting the two $\mathcal{N}=1$ superfield actions and eliminating the auxiliary fields. The explicit form of the actions in components together with the computational conventions can  be found in appendix \ref{appendix:conventions}.  

In this context we consider a Maldacena--Wilson operator common to $\mathcal{N}=2$ SCQCD and $\mathcal{N}=4$ SYM theories  
\begin{equation}\label{WL}
W = \frac{1}{N} \textrm{Tr} \,\mathcal{P} \,e^{- ig\int_C d\t \mathcal{L}(\t)}  
\end{equation}
with Euclidean connection 
\begin{equation}
\label{connection}
\mathcal{L}(\t) = \dot{x}^{\mu}A_{\mu}+\frac{i}{\sqrt{2}} \, | \dot{x} | \, (\phi+\bar{\phi}) 
\end{equation}
where $\phi, \bar{\phi}$ are the adjoint scalars entering the $\mathcal{N}=2$ vector multiplet shared by the two theories.  

Even if our computation is entirely done in the component formalism, it is worth to mention that in $\mathcal{N}=1$  superspace some effective rules to evaluate diagrammatic difference $\langle O\rangle_{\mathcal{N}=2} - \langle O\rangle_{\mathcal{N}=4}$ have been derived \cite{Pomoni:2011jj} and later formalized  \cite{Pomoni:2013poa}  in the context of the   calculation of the $SU(2,1|2)$  spin chain Hamiltonian  of  $\mathcal{N}=2$ SCQCD. In that case it was shown that the only source of diagrams potentially contributing to the difference is given by graphs containing chiral loops cut by an adjoint line (either vector or chiral). This rule was found later to be valid also in the context of the computation of the adjoint scattering amplitudes of $\mathcal{N}=2$ SCQCD \cite{Leoni:2015zxa}.  In particular, topologies containing ``empty" chiral loops are constrained to produce the same result for the two models. In fact, for such type of diagrams computing the difference is only a matter of counting the number of possible realizations of the loop in terms of adjoint  and/or fundamental superfields. As a consequence of the condition $N_f=2N$, the two models turn out to give the same result. 

One might wonder whether similar rules survive when reducing the theory to components and if they  can be easily applied to the computation of the cusp anomalous dimension. As we are going to show in the rest of the paper, this turns out to be the case for diagrams involving only minimal gauge matter-couplings up to three loops, due to the fact that the actions in components display the same flavour structure of their   $\mathcal{N}=1$ superspace versions.
Of course, possible complications in taking too seriously the  parallel with the superfield  rules may arise when considering higher order diagrams involving superpotential vertices. In this case we expect the component diagramatics to follow different rules with respect to the superspace version.

\section{The Matrix Model result}\label{MM}

In order to prove identity \eqref{B2} we begin by recalling the evaluation of its right hand side, where $ \langle W_b\rangle$ is the 1/2 BPS circular Wilson loop of the form (\ref{WL}, \ref{connection}) defined on the maximal latitude or the maximal longitudinal circles of the ellipsoid \cite{Fiol:2015spa}
\begin{equation}
 x_0^2+\frac{x_1^2+x_2^2}{l^2}+\frac{x_3^2+x_4^2}{\widetilde{l}^2}=1
\end{equation}
 
Applying localization techniques, in $\mathcal{N}=4$ SYM theory the vev of this operator is computed exactly by the following matrix model  \cite{Erickson:2000af, Drukker:2000rr, Pestun:2007rz}    
\begin{equation} 
 \langle W_b\rangle = \frac{\int da\;\mathrm{tr}(e^{-2\pi ba})\,e^{-\frac{8\pi^2N}{\lambda}\mathrm{tr}(a^2)}}{\int da\;e^{-\frac{8\pi^2 N}{\lambda}\mathrm{tr}(a^2)}}+\mathcal{O}\left((b-1)^2\right)
\label{Wb}
 \end{equation}
and turns out to be a function of the squashing parameter $b=(l/\widetilde{l})^2$.
From this expression the Bremsstrahlung function can be easily computed by using identity \eqref{B4b}.

Similarly, in $\mathcal{N}=2$ SCQCD it is given by 
\cite{Pestun:2007rz, Alday:2009fs, Passerini:2011fe, Hama:2012bg, Fucito:2015ofa, Fiol:2015mrp} (for a review, see \cite{Okuda:2014fja})
\begin{eqnarray}
 \langle W_b\rangle=  
 \frac{\int da \mathrm{Tr}\,e^{-2\pi ba}\,e^{-\frac{8\pi^2}{g^2}\mathrm{Tr}(a^2)}Z_{1-loop}(a,b)|Z_{inst}(a,b)|^2}{\int da\,e^{-\frac{8\pi^2}{g^2}\mathrm{Tr}(a^2)}Z_{1-loop}(a,b)|Z_{inst}(a,b)|^2} 
 \label{Wb2}
 \end{eqnarray}
 
According to conjecture \eqref{B2} the only terms in the matrix model which can contribute to $B$ are the ones linear in $(b-1)$. Since the classical, one-loop and instanton contributions start deviating from their $S^4$ counterparts only at second order in $(b − 1)$, it follows that $\langle W_b\rangle$ in \eqref{Wb2} can be computed using the one--loop determinant and instanton factors of the round $S^4$ matrix model \cite{Fiol:2015spa}.\\ 
\\
Assuming prescription \eqref{B2} to be true for any $N$ and expanding the two matrix models (\ref{Wb}, \ref{Wb2}) up three loops, we obtain the general prediction for the difference of the Bremsstrahlung function in the two theories
\begin{equation}\label{prediction}
 B_{\mathcal{N}=2} - B_{\mathcal{N}=4} = - \frac{3\zeta(3)}{1024 \pi^6}\, \frac{(N^2-1)(N^2+1)}{N} \, g^6 +\mathcal{O}(g^8)
\end{equation}
For $N=2$ this expression reduces to 
\begin{equation}
 B_{\mathcal{N}=2}-B_{\mathcal{N}=4}= -  \frac{45}{2048\pi^6}\, \zeta(3) \, g^6 +\mathcal{O}(g^8)
 \label{7}
\end{equation}
which has been already checked in \cite{Fiol:2015spa} against a three--loop perturbative calculation.
In the next section we generalize the proof of eq. \eqref{prediction} to any finite $N$.

\section{The perturbative result  }
\label{section3}
In order to check eq. \eqref{prediction} we perform a perturbative three--loop calculation of its left hand side along the lines described in the introduction, that is by extracting the difference of the two bremsstrahlung functions from the small angle limit of the difference of the corresponding cusp anomalous dimensions, $\Gamma_{\mathcal{N}=2}-\Gamma_{\mathcal{N}=4}$.  

To this end, in Euclidean space we consider an operator of the form \eqref{WL} where the contour $C$ is made by two infinite straight lines parametrized as
\begin{eqnarray}
&  x^{\mu}(\tau_1)=v_1^{\mu}\tau_1 &\;  \quad \qquad 0<\tau_1<\infty \nonumber \\
&  x^{\mu}(\tau_2)=v_2^{\mu}\tau_2 & \qquad -\infty<\tau_2<0
\end{eqnarray}
The two lines form an angle $\varphi$, such that $\cos\varphi=v_1\cdot v_2$ and $|v_1|  = |v_2| = 1$. 

We also allow for two different scalar couplings on the two lines of the contour, characterised by a relative internal angle $\theta$. Precisely, on the two rays we choose
\begin{align}
\mathcal{L}_1(\t) & = v_1^{\mu}A_{\mu} +\frac{ i}{\sqrt{2}} (\phi \, e^{i \theta/2}+\bar{\phi} \, e^{-i \theta/2}) \\
\mathcal{L}_2(\t) & = v_2^{\mu}A_{\mu} + \frac{ i}{\sqrt{2}} (\phi \, e^{-i \theta/2}+\bar{\phi} \, e^{i \theta/2}) 
\end{align}
 
As a first step we have to evaluate $W_{\mathcal{N}=2}-W_{\mathcal{N}=4}$. At order $\mathcal{O}(g^2)$ the only diagrams are the single gluon and single adjoint scalar exchanges, for which the result is the same in both theories. At this order the difference is therefore zero. This property extends to all the diagrams built with tree level $n$--point functions inserted into the Wilson line, since in this case contributions from the hypermultiplets do not appear. The next order is $\mathcal{O}(g^4)$, where the only non--tree diagrams are the exchange of one--loop corrected propagators. However, it has been shown that in the difference they still cancel since the contribution from a loop of $2N$ fundamental fields is the same as the one from the loop of one adjoint field \cite{Andree:2010na}. 

The first non--trivial contribution starts at $\mathcal{O}(g^6)$ where the contributing diagrams correspond to the insertion of two--loop corrected gauge/scalar propagators and one--loop corrected cubic vertices. Here we analyze them separately, and postpone the evaluation of the corresponding integrals to section \ref{section4}.

\subsection{Two--loop propagator diagrams}

\begin{figure} [H]
\centering
{\includegraphics[width =10cm]{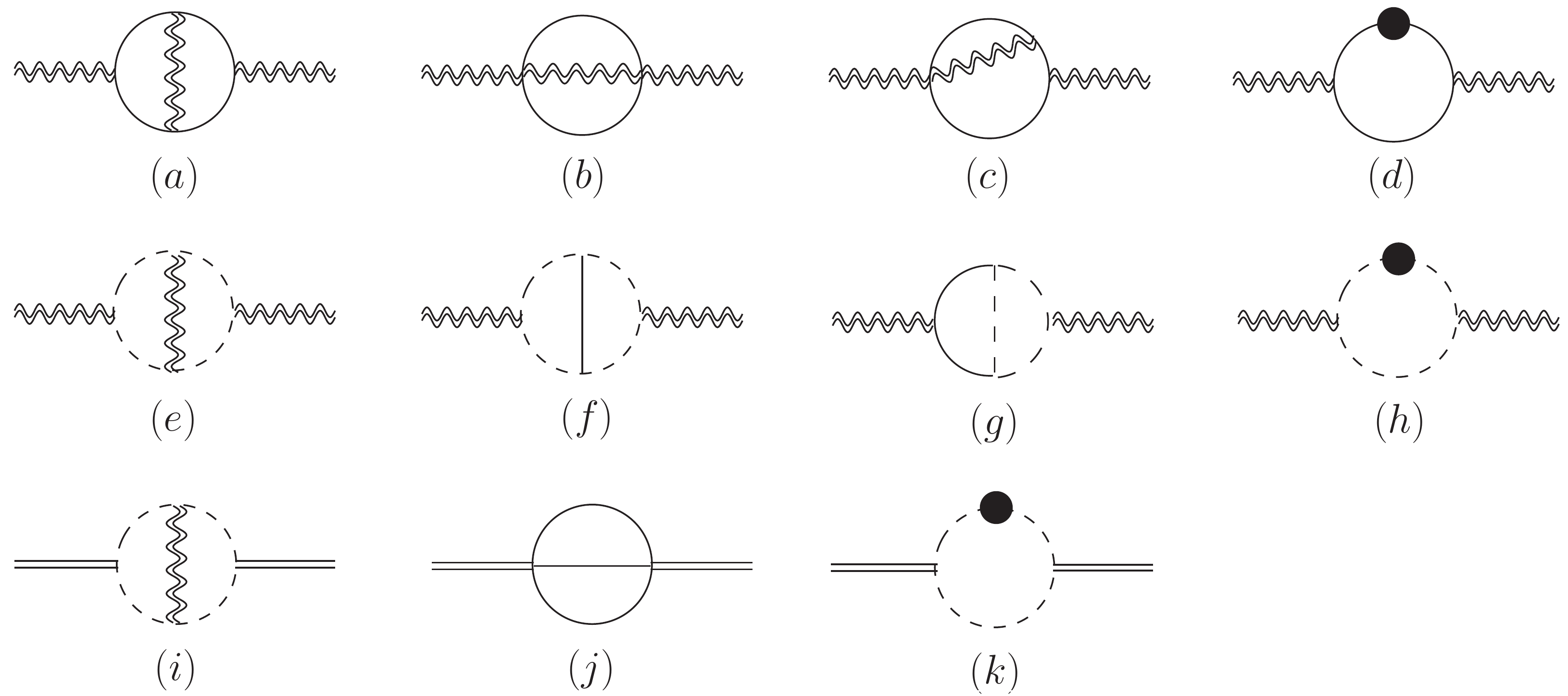}}
\caption{Diagram topologies that contribute to the difference of the  $\mathcal{N}=2$ and $\mathcal{N}=4$ propagators at two loops.}
\label{diag}
\end{figure}

We begin by considering the diagrams with  two--loop corrections to the vector and adjoint scalar propagators. Taking the difference between the $\mathcal{N}=2$ and $\mathcal{N}=4$ propagators, the diagram topologies which survive  are the ones listed in figure \ref{diag}. Here we neglect topologies that would produce vanishing cusp integrals. 

\noindent
For simplicity we are not depicting the insertion of the diagrams into the Wilson loop contour. 
We use double lines to represent fields in the adjoint representation, whereas we use simple lines to represent fields that can be either in the adjoint or in the fundamental representation.  Each topology in figure \ref{diag} then corresponds to the collection of all possible diagrams of that kind that can be realized in terms of both adjoint and fundamental fields of the two models.
For instance, in figure 1$(a)$ the simple solid loop stands generically for one of the following realizations: In 
$\mathcal{N}=4$ SYM it indicates any of the three adjoint scalar fields $\phi^I$, $\, I=1, 2,3$, whereas in $\mathcal{N}=2$ SCQCD it corresponds to either the adjoint scalar $\phi$ or one of the two fundamental sets of fields  $q_I, \bar{\tilde{q}}_I$ with $I=1,\dots,2N$.  The same happens for diagrams $(b),(c), (d)$. 
For diagrams $(e), (h), (i), (k)$ involving a simple fermionic loop we have a parallel counting, this time in terms of the adjoint fermion fields $\psi$ and the fundamentals $\lambda, \tilde{\l}$. 

We see that, excluding  diagrams $(f), (g), (j)$, we are only  dealing with  minimal gauge--matter couplings, so that the superfield difference selection rules of \cite{Pomoni:2013poa} still hold and we are left only with diagrams with matter loops cut by an adjoint line. 
Instead, diagrams $(f), (g), (j)$ involve interaction vertices from the potential. Consequently, the list of possible field realizations cannot  exactly parallel the superfield counting anymore. For instance, diagram $(g)$  produces non-vanishing contributions to the difference which include the gaugino field $\eta$, while diagram $(j)$ requires a careful analysis of all possible flavour realizations stemming from the quartic vertices.  

It is interesting to note that diagrams $(d), (h), (k)$, which are generated by the 1--loop corrected fermion and scalar propagators, do not have a correspondent in $\mathcal{N}=1$ superspace. In fact, in a $\mathcal{N}=1$ superspace setup the 1-loop corrections to the superfield propagators are exactly vanishing for both  $\mathcal{N}=4$ SYM and $\mathcal{N}=2$ SCQCD.  However, in the component formulation this is no longer the case and the one-loop corrections turn out to be divergent. This is not in contradiction with conformal invariance and can be interpreted as a consequence of working in the susy-breaking Wess--Zumino gauge \cite{Kovacs:1999rd}.

\subsection{One--loop three--point vertex diagrams}

\begin{figure} [H] 
\centering
{\includegraphics[scale=0.3]{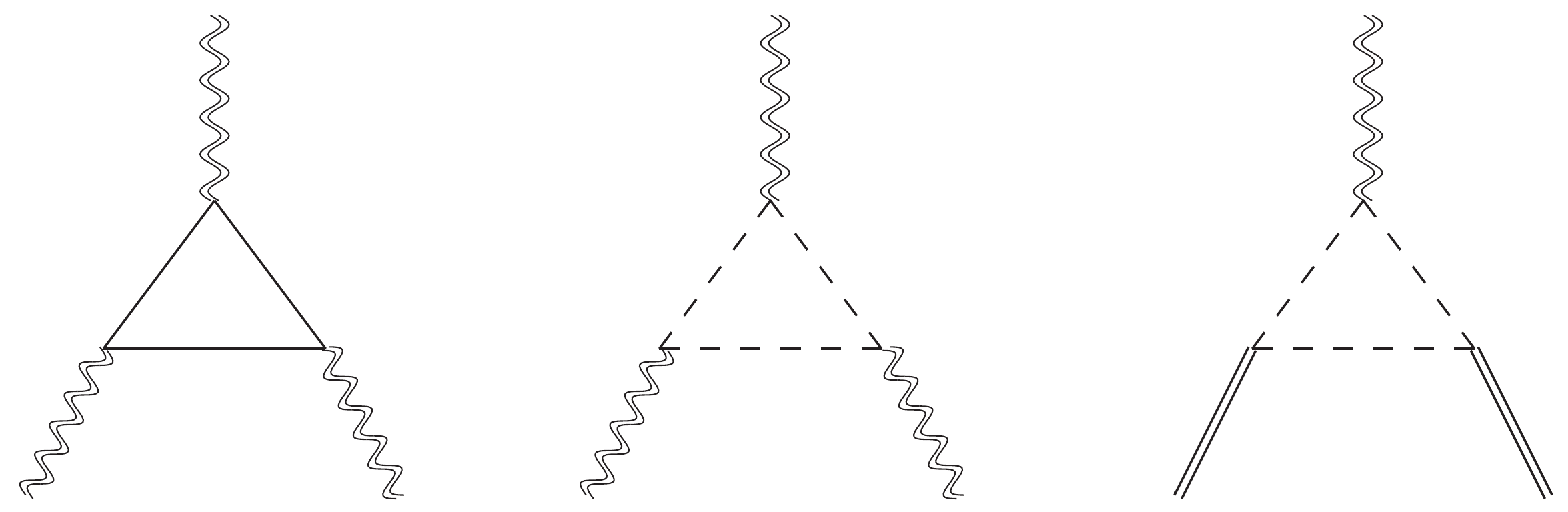}}
\caption{One--loop corrections to the three--point vertices that potentially contribute to the difference.}
\label{diag2} 
\end{figure}

In principle, other contributions at order $g^6$ may come from the insertion into the cusp  of the one--loop corrections to the three--point vertices \cite{Andree:2010na, Fiol:2015spa}. The diagram topologies potentially contributing to the difference between the $\mathcal{N}=2$ and $\mathcal{N}=4$ three--point vertices are depicted in figure \ref{diag2}. Again we neglect topologies that would produce vanishing cusp integrals. 
In  \cite{Fiol:2015spa, Andree:2010na} it has been proved that in the $SU(2)$ case the contribution to the difference $W_{\mathcal{N}=2}-W_{\mathcal{N}=4}$ is vanishing at the conformal point, due to the fact that for algebraic reasons the result from two adjoint scalars running into the loop is identical to the result from $2N$ fundamentals. However, as argued in \cite{Andree:2010na}, the result cannot be immediately generalized to $SU(N)$, since for $N>2$ extra contributions from the adjoint scalar loop may arise, which are proportional to the symmetric structure constants $d_{abc}$ (see eq. \eqref{coloridentities}). 

Here we perform a detailed analysis of these diagrams and prove that for symmetry reasons contributions proportional to $d_{abc}$ can never appear. Therefore, we conclude that diagrams in figure \ref{diag2} never contribute to the difference $W_{\mathcal{N}=2}-W_{\mathcal{N}=4}$, for any $SU(N)$  gauge group, so generalizing the result of \cite{Fiol:2015spa, Andree:2010na}.   

 We illustrate how the cancelation works by focusing on an explicit example, that is the  first  vertex topology in figure \ref{diag2} that corresponds to the scalar loop corrections to the three--gluon vertex.  

 In the $\mathcal{N}=4$ SYM case  we have  the three adjoint scalars $\phi^I$, $\, I=1,2,3$ running into the loop. Using Feynman rules in appendix \ref{appendix:conventions} the corresponding expression reads 
\begin{align} \label{3pt}  
V_{\mathcal{N}=4}  =  3 \, N \,g^3\,  \textrm{Tr}\big(T^a [T^b, T^c]\big) \int d^{d}z_{1/2/3}  \, f^{\mu \nu \rho}(z_1, z_2, z_3)  
\times  A^a_{\mu}(z_1)A^b_{\nu}(z_2)A^c_{\rho}(z_3)  
\end{align}
where the factor 3 stems from the sum over all possible flavour loops, the two terms building up the color trace commutator correspond to the two possible orientations of the adjoint loop cycles, and $ f^{\mu \nu \rho}$ is a function of the vertex points $z_{1/2/3}$ that can be expressed in momentum space as 
\begin{align} \label{integral}  
 f^{\mu \nu \rho}(z_1, z_2, z_3)   =  \!\int \!\frac{d^{d}(q+k_{2})}{(2\pi)^{d}}& \!\int \!\frac{d^{d}(q-k_{1})}{(2\pi)^{d}} \!\int\!\frac{d^{d} q}{(2\pi)^{d}} \, e^{i q(z_1-z_2)}e^{i(q+k_2)(z_2-z_3)}e^{i(q-k_1)(z_3-z_1)}  \nonumber    \\
&  \times \,\, \frac{(2q-k_1)^{\mu}(2q +k_2)^{\nu}(2q+k_2-k_1)^{\rho}}{ q^2 (q-k_1)^2(q+k_2)^2} 
\end{align}
\vspace{0.2cm} 
In $\mathcal{N}=2$ SCQCD the same kind of diagram topology can be constructed using either the single adjoint scalar $\phi$ or the two fundamental sets of fields $q_I, \bar{\tilde{q}}_I$ with $I=1,\dots, 2N$. The adjoint loop will give exactly the same result as in \eqref{3pt}, without the  factor 3. Instead,  the two sets of fundamental loops yield
\begin{align} \label{3ptN2}  
 V^{fund}_{\mathcal{N}=2}   =       2 \times 2N \, g^3\, \textrm{Tr}\big(T^a T^b T^c\big) \, f^{\mu \nu \rho}(z_1, z_2, z_3) 
  \times  A^a_{\mu}(z_1)A^b_{\nu}(z_2)A^c_{\rho}(z_3) 
\end{align}
where now we have a single possible color orientation and the integral is still given in \eqref{integral}. 

Taking the difference we obtain 
\begin{align} \label{3ptdiff}  
&\!\! \!  V_{\mathcal{N}=2}- V_{\mathcal{N}=4} = \! \big\{  4 N  \textrm{Tr}\big(T^a T^b T^c\big)  - 2  N  \textrm{Tr}\big(T^a [T^b, T^c]\big) \big\} \nonumber \\ & \hspace{0.01cm}\times\,g^3 \, \int d^{d}z_{1/2/3}  \, f^{\mu \nu \rho}(z_1, z_2, z_3) \,   \times A^a_{\mu}(z_1)A^b_{\nu}(z_2)A^c_{\rho}(z_3) 
\end{align}
where for $SU(2)$ the color structure inside the bracket is identically vanishing, whereas for $SU(N)$ it is nothing but the totally symmetric $d^{abc}$ tensor (see eq. \eqref{identity}). 

It is now easy  to see that, independently of the gauge group, this expression always vanishes. In fact, the string $d_{abc} A^a_{\mu}(z_1)A^b_{\nu}(z_2)A^c_{\rho}(z_3) $ is symmetric under the exchange of any pair of gauge fields, but it is contracted with $f^{\mu \nu \rho}$ which is antisymmetric under any exchange
\begin{equation}
f^{\mu \nu \rho}(z_1, z_2, z_3) = - f^{\nu \mu\rho}(z_2, z_1, z_3) \qquad {\rm etc...}
\end{equation}
 
An alternative reasoning goes as follows. Independently of the gauge group, once we insert the vertex correction \eqref{3ptdiff} into the cusp contour, for symmetry reasons the two color trace structures of the commutator term $\textrm{Tr}\big(T^a [T^b ,T^c] \big)$ sum up to $2\textrm{Tr}\big(T^a T^b T^c\big)$, so that the difference in \eqref{3ptdiff} vanishes identically. This can be loosely summarized stating that each adjoint empty loop counts as twice a fundamental loop contribution, thus producing a vanishing counting. 

It is easy to realize that similar symmetry arguments hold for all the other topologies in figure \ref{diag2}.  We then conclude that, against previous expectations, there are no contributions to
$W_{{\cal N}=2} - W_{{\cal N}=4}$ coming from one--loop three--point vertices, for generic $SU(N)$ gauge group.

\section{The computation of the diagrams and HQET procedure}
\label{section4}
According to the previous discussion, the only non--trivial contributions to the difference $W_{\mathcal{N}=2}-W_{\mathcal{N}=4}$ come from the insertion of diagrams in figure \ref{diag}. In this section we focus on the evaluation of the corresponding loop integrals. 

We can focus only on insertions which connect the two lines of the cusped Wilson loop (1PI diagrams in the HQET context) since the ones where the two insertion points lie both on the same ray can be factorized out and do not contribute to the evaluation of B \cite{Bianchi:2017svd}. 

The most efficient way to compute the corresponding loop integrals is the so--called HQET method  \cite{Grozin:2015kna}. Working in momentum space, it consists in integrating first on the contour parameters with a proper prescription for regularizing boundary divergences. This reduces the integrals to ordinary massive momentum integrals, which can be written as linear combinations of known Master Integrals by applying integrations by parts.

The full list of results for diagrams of figure \ref{diag} can be found in appendix \ref{appendix:partial}. Here we briefly illustrate the procedure by computing for instance the integral corresponding to diagram $(e)$. In the $\mathcal{N}=4$ SYM case the fermionic loop, represented in our notation with a simple dashed line, can be constructed with any of the three adjoint fermions  $\psi^I$, with $\, I=1, 2,3$.
In the $\mathcal{N}=2$ SCQCD case, instead, the loop can be realized either with the adjoint fermion $\psi$ or with one of the two sets of fundamental fermions $\lambda_I, \tilde{\l}^I$, with $\, I=1, ... ,2N$. Taking the difference of $\mathcal{N}=2$ and $\mathcal{N}=4$ propagators and inserting it in the Wilson line, the corresponding integral reads (we neglect a factor  $\frac{g^6(N^2-1)(N^2+1)}{2 N}$ )
\begin{eqnarray}
 & &  I^{(e)} =  -   \int_{0}^{\infty}d\tau_1\int_{-\infty}^{0}d\tau_2 \,  v_1^\mu\,v_2^\nu\, \text{tr}(\sigma^{\mu}\sigma^{\rho}\sigma^{\xi}\sigma^{\tau}\sigma^{\nu}\sigma^{\sigma}\sigma_{\xi}\sigma^{\eta}) \nonumber \\
 &  & \int\dfrac{d^{d}k_{1/2/3}}{(2\pi)^{3d}}\,e^{ik_3\cdot(x_1-x_2)}  \,\dfrac{(k_1-k_3)_{\rho}(k_1)_{\eta}(k_2-k_3)_{\tau}(k_2)_{\sigma}}{k_1^2\,k_2^2\,k_3^4\,(k_1-k_2)^2\,(k_1-k_3)^2\,(k_2-k_3)^2} 
\label{diag-e-loop}
 \end{eqnarray}
where we work in $d = 4 - 2\e$ dimensions and we have defined $x_1^\mu \equiv v_1^\mu \tau_1$, $x_2^\mu \equiv v_2^\mu \tau_2$.

Now the trick consists in changing the order of contour and momentum integrals and perform first the contour ones. This amounts to first compute 
\begin{eqnarray}
& & \int_{0}^{\infty}d\tau_1\;e^{ik_3\cdot v_1\tau_1}e^{\delta\tau_1}\; \int_{-\infty}^{0}d\tau_2\;e^{-ik_3\cdot v_2\tau_2}e^{-\delta\tau_2}  \nonumber \\
  &=&  \frac{1}{(ik_3\cdot v_1+\delta)}\frac{1}{(ik_3\cdot v_2+\delta)}\label{contourreg}
\end{eqnarray}
where, following the prescription of \cite{Grozin:2015kna}, a damping factor $e^{\delta\tau}$ with $\delta<0$ has been introduced for each contour integral in order to make them well defined at infinity. 

Since the final result is expected to be independent of the IR regulator $\delta$, we conveniently choose $\delta=-1/2$.  Absorbing the $i$ factor in a redefinition of the velocity, $v=i\,\widetilde{v}$, we are left with
\begin{eqnarray}
 I^{(e)} =  4  \int\dfrac{d^{d}k_{1/2/3}}{(2\pi)^{3d}}\dfrac{\text{tr}(\sigma^{\mu}\sigma^{\rho}\sigma^{\xi}\sigma^{\tau}\sigma^{\nu}\sigma^{\sigma}\sigma_{\xi}\sigma^{\eta})}{(1+2k_3.\widetilde{v}_1)(1+2k_3.\widetilde{v}_2)} \times \dfrac{\tilde{v}_{1\mu} \, \tilde{v}_{2\nu} \, (k_1-k_3)_{\rho}(k_1)_{\eta}(k_2-k_3)_{\tau}(k_2)_{\sigma}}{k_1^2k_2^2k_3^4(k_1-k_2)^2(k_1-k_3)^2(k_2-k_3)^2} \nonumber \\
\label{diag-e-loop}
 \end{eqnarray}

Now, using the $\sigma$-matrix algebra we can reduce the numerator to a linear combination of scalar products of momenta and external velocities which can be written in terms of inverse propagators. Therefore,  we end up with a sum of momentum integrals of the form \ref{G-int}. These integrals are not all independent and, using integration by parts, performed with the Mathematica package FIRE \cite{Smirnov:2013dia, Smirnov:2014hma, Smirnov:2008iw},  they can be expressed in terms of a finite set of Master Integrals \cite{Grozin:2015kna, Grozin:2014hna}. For our example, after the FIRE reduction we obtain
\begin{eqnarray}
& I^{(e)} = \left[\left(\frac{32 (3 d-7)(-480+964 d-796 d^2+335 d^3-71 d^4+6 d^5)}{ (d-5) (d-4)^3 (d-3) (d-1)}\right. \right. \nonumber \\
& \left. \left. +\frac{32 (3 d-7)(-4736+8360 d-5494 d^2+1663 d^3-222 d^4+9 d^5)\cos\varphi }{3 (d-5) (d-4)^3 (d-3) (d-1)}  \right.\right. \nonumber \\
& \left. \left. +\frac{32 (3 d-7)(-2720+4736 d-3196 d^2+1036 d^3-159 d^4+9 d^5) \cos^2 \varphi }{3 (d-5) (d-4)^3 (d-3) (d-1)}  \right) \times\, I_1 \right. \nonumber \\
& \left. -\left(\frac{16 (1+\cos\varphi)^2 (80-54 d+9 d^2)(96-140 d+81 d^2-21 d^3+2 d^4)}{ (d-5) (d-4)^3 (d-3) (d-1)} \right.\right. \nonumber \\
& \left. \left. -\frac{16 (1+\cos\varphi)^2 (80-54 d+9 d^2)(272-392 d+202 d^2-43 d^3+3 d^4)\cos\varphi}{3 (d-5) (d-4)^3 (d-3) (d-1)}\right) \times\, I_2 \right. \nonumber \\
& \left. -\frac{12 (d-3) (10-\cos\varphi (d-8)-3 d) (8-5 d+d^2)}{(d-5) (d-4)^2 (d-1)} \times \, I_3 \right. \nonumber\\
& \left. -\frac{2 (1+\cos\varphi) (\cos\varphi (d-8)-3 (d-4)) (80-74 d+25 d^2-3 d^3)}{(d-5) (d-4)^2 (d-1)} \times\, I_4\right]
\end{eqnarray}
where the Master Integrals $I_i$ are defined in Appendix \ref{appendix:partial}. 

This technique is known as ``Heavy Quark Effective Theory'' (HQET) due to its relation with the theory of scattering of heavy particles. The propagator-like integrals that we obtain with the method described above formally coincide with the integrals describing the propagation of heavy quarks. The direction $v^{\mu}$ of the Wilson line is the velocity of the quark, whereas the damping factor $\delta$ corresponds to the introduction of a residual energy for the particle. In the presence of a cusp, the Bremsstrahlung function controls the energy radiated by the heavy particle undertaking a transition from a velocity $v_1$ to $v_2$ in an infinitesimal angle $\varphi$. \\

Since we are eventually interested in computing the cusp anomalous dimension that in dimensional regularization can be read from the $1/\epsilon$ pole of $\log \langle W \rangle$, it is convenient to expand the master integrals in powers of $\epsilon$. Defining the new variable $x=e^{i\varphi}$, where $\varphi$ is the geometric angle of the cusp, for the $I^{(e)}$ we find 
\begin{eqnarray}
& I^{(e)} =     \dfrac{1}{\epsilon^3} \, \frac{2 \left(1-x^2-2(1+x^2) \log[x]\right)}{9 \left(x^2 -1\right)}   \nonumber \\
& +  \dfrac{2}{\epsilon^2} \, \left( \frac{4-\pi ^2-(4+\pi^2) x^2+3 (1+x^2) \log[x]^2}{9 \left(x^2-1\right)}  - \frac{2 \log[x] \left(5+x (5 x-3)+6 (1+x^2) \log[1+x]\right)+12 (1+x^2) \text{Li}_2[-x]}{9 \left(x^2-1\right)} \right)   \nonumber\\
&+ \dfrac{1}{\epsilon} \, \left( \frac{80-7 \pi ^2+12 \pi ^2 x-(80+33 \pi ^2) x^2}{18 \left(x^2-1\right)} -  \frac{\log[x]   \left(101-(48-101 x) x+7 \pi ^2 (1+x^2)\right)}{9 \left(x^2-1\right)}  \right. \nonumber \\
&+ \left.\frac{6 \log^2[x] \left(5-(3-5 x) x-(1+x^2) \log[x]\right)}{9 \left(x^2-1\right)} - \frac{12 \left(3 \pi ^2 (1+x^2)+2 (5+x (-3+5 x)) \log[x]-3 (1+x^2) \log[x]^2\right) \log[1+x]}{9 \left(x^2-1\right)} \right.\nonumber \\
&+ \left. \frac{144(1+x^2) (\log[-x]-\log[x]) \log[1+x]^2}{18\left(x^2-1\right)} - \frac{48 (5+x (-3+5 x)) \text{Li}_2[-x]-144 (1+x^2) \text{Li}_3[-x]-288 (1+x^2) \text{Li}_3[1+x]}{18\left(x^2-1\right)} \right. \nonumber \\
& \left. +\frac{96 \left(-2-x^2+\log[x]+x^2 \log[x]\right) \zeta[3]}{18 (x^2-1)}\right)  + \mathcal{O}(\epsilon^0)
\end{eqnarray}
The expansions of the integrals corresponding to the rest of the diagrams in figure \ref{diag} are listed in appendix \ref{appendix:partial}. We note that the expansions may contain higher order poles in $\e$, up to $1/\e^3$.

\section{The result}
\label{section5}

Applying the HQET procedure to every single diagram and summing the results for the integrals as listed in appendix \ref{appendix:partial}, we can distinguish the contribution coming from the insertion of diagrams $(a) - (h)$ (insertion of a gauge propagator)
\begin{eqnarray}
 &  \big[ \langle W_{\mathcal{N}=2}\rangle-\langle W_{\mathcal{N}=4}\rangle \big]\bigg|_{gauge}^{(3L)} = g^6\, \dfrac{(N^2-1)(N^2+1)}{2048 \,\pi^6 N} \zeta(3) \, \dfrac{-1+x^2+2(1+x^2)\log{x} }{(x^2 -1)\epsilon} \qquad
\end{eqnarray}
from the contribution arising from diagrams $(i) - (k)$  (insertion of an adjoint scalar propagator)  
\begin{eqnarray}
\big[ \langle W_{\mathcal{N}=2}\rangle-\langle W_{\mathcal{N}=4}\rangle \big]\bigg|_{scalar}^{(3L)} = - g^6 \, \frac{ (N^2-1)(N^2+1)}{2048 \, \pi^6N}  \zeta(3) \, \cos\theta \, \frac{4x\log{x} \,  }{(x^2-1)\epsilon}
\end{eqnarray}
It is remarkable that, although individually the integrals corresponding to the various topologies in figure \ref{diag} exhibit up to cubic poles in $\epsilon$, in the sum they all cancel and only a simple pole survives.  This has a simple physical explanation and represents a non--trivial consistency check of our calculation.  In fact,  
 according to equation  \eqref{CAD}, which in dimensional regularization reads $\langle W \rangle \sim \exp{(\Gamma(g^2)/\epsilon)}$, higher order $\epsilon$-poles  in the Wilson loop expansion only come from the exponentiation of $\frac{\Gamma(g^2)}{\epsilon}$.  Since the difference $\langle W_{\mathcal{N}=2}\rangle-\langle W_{\mathcal{N}=4}\rangle$ is identically vanishing up to two loops, at three loops we expect to find only simple poles. Taking into account that the exponentiation works also when we turn off the scalar coupling,  both the gauge and the scalar contributions   have to display the higher order poles cancellation, independently.

Now, summing the two contributions and defining $\xi=\dfrac{1+x^2-2x\cos\theta}{1-x^2}$,  we obtain
\begin{eqnarray}
 \langle W_{\mathcal{N}=2}\rangle-\langle W_{\mathcal{N}=4}\rangle \, = \, g^6 \, \frac{\zeta(3)}{2048 \, \pi^6}\frac{(N^2 − 1)(N^2 + 1)}{N} \times \left(1 - 2 \xi \log{x} \right) \, \frac{1}{\epsilon} + {\cal O}(g^8) \nonumber \\
\end{eqnarray}

The presence of the IR regulator $e^{\delta\tau}$ inside the contour integrals breaks gauge invariance. As a consequence, gauge--dependent spurious divergences survive, which need to be eliminated prior computing the cusp anomalous dimension. 
 As explained in details in \cite{KORCHEMSKY1987342, Bianchi:2017svd}, this can be done by introducing a multiplicative renormalization constant $Z_{\text open}$, which in practice corresponds to remove the value at $\varphi=\theta=0$
\begin{eqnarray}\label{IRremoving}
 \langle\widetilde{W}(\varphi,\theta)\rangle \equiv Z^{-1}_{open}\langle W(\varphi,\theta)\rangle=\frac{\langle W(\varphi,\theta)\rangle}{\langle W(0,0)\rangle} 
\end{eqnarray}
We then obtain the IR--divergence free difference, which reads
\begin{eqnarray}\label{IRremoving2}
 \langle \widetilde{W}_{\mathcal{N}=2}\rangle-\langle \widetilde{W}_{\mathcal{N}=4}\rangle
\, \,= \, \, - g^6 \, \frac{\zeta(3)}{1024 \, \pi^6}\frac{(N^2 − 1)(N^2 + 1)}{N} \times \xi \log{x} \, \times \frac{1}{\epsilon} + {\cal O}(g^8) \non \\
\end{eqnarray}
Recalling that in dimensional regularization with $d= 4 - 2\e$ we need to rescale $g \to g \mu^{-\epsilon}$ where $\mu$ is a mass scale, and using definition \eqref{CAD} we can easily read the difference of the two cusp anomalous dimensions from the aforementioned $1/\epsilon$ pole, obtaining 
\begin{equation}\label{Gfinal}
 \Gamma_{\mathcal{N}=2}-\Gamma_{\mathcal{N}=4} \, \,  = \, \, g^6 \, \frac{3\zeta(3)}{512 \, \pi^6}\frac{(N^2 − 1)(N^2 + 1)}{N} \, \xi \log{x}  + {\cal O}(g^8)
 \end{equation}
This equation represents the most complete result for the three--loop deviation of $\Gamma_{{\cal N}=2}$ from $\Gamma_{{\cal N}=4}$. In particular, it is valid for any finite $\theta, \varphi$ and $N$.

Remarkably, we find that at $\theta=\pm\varphi$ eq. \eqref{Gfinal} vanishes, suggesting that at these points the cusped Wilson loop of $\mathcal{N}=2$  SCQCD  might become 1/2 BPS as in the ${\mathcal{N}=4}$ SYM case.
  
Now, re-expressing $\xi$ and $x$ in terms of the original $\theta, \varphi$ variables and taking the small angle limit, $2 \xi \log{x}  {\underset{\varphi, \th \ll 1}{\sim}} (\varphi^2 - \theta^2)$,  we obtain the difference of the corresponding Bremsstrahlung functions
\vspace{0.3cm}
\begin{equation}
\label{Bfinal}
 B_{\mathcal{N}=2} -B_{\mathcal{N}=4} = -g^6 \, \frac{3\zeta(3)}{1024 \,\pi^6}\frac{(N^2-1)(N^2+1)}{N}+\mathcal{O}(g^8)
\end{equation}
\\
This result remarkably coincides with prediction \eqref{prediction} from the matrix model. We have then found confirmation at three loops that conjecture \eqref{B2} proposed in  \cite{Fiol:2015spa} is valid for any $SU(N)$ gauge group.

If we insert the known value of $B_{\mathcal{N}=4} $ \cite{Correa:2012nk}, in the large $N$ limit we  find 
\begin{eqnarray}
\label{Bfinalcomplete}
& B_{\mathcal{N}=2} = \dfrac{g^2 N}{16\pi^2}-\dfrac{g^4 N^2}{384\pi^2}+\dfrac{g^6N^3}{512\pi^2}\left(\dfrac{1}{12}\,-  \dfrac{3\zeta(3)}{2 \, \pi^4}\right)+\mathcal{O}(g^8)
\end{eqnarray}

\subsection{Light--like cusp}

Given our previous results, it is interesting to study the limit of large Minkowskian angles. To this end we substitute $\varphi=i\varphi_M$, that is $x=e^{-\varphi_M}$, and send $x\to0$. In this limit the cusp anomalous dimension behaves linearly in the angle
\begin{equation}
 \Gamma_{\text{cusp}}(g^2,N,\varphi){\underset{\varphi\to\infty}{\sim}}\,K(g^2,N)\varphi_M+\mathcal{O}(\varphi_M^0)
\end{equation}
The function $K(g^2,N)$ is called the light-like cusp anomalous dimension. 

Using the large--$N$ exact results for the $\mathcal{N}=4$ SYM case previously found in the literature (\cite{Correa:2012nk} \cite{Grozin:2014hna}), for $\mathcal{N}=2$ SCQCD we obtain  
\begin{equation}\label{lightlikecusp}
K_{\mathcal{N}=2}(g^2,N)=\dfrac{g^2N}{8\pi^2}-\dfrac{g^4N^2}{384\pi^2}+\dfrac{g^6N^3}{512\pi^2}\left(\dfrac{11}{180}-\dfrac{3\zeta(3)}{\pi^4}\right) +\mathcal{O}(g^8)
\end{equation}
The light--like cusp can be used to check an interesting universal behaviour of the cusp anomalous dimension that was found to hold up to three loops in QCD and in Yang--Mills theories with only adjoint matter \cite{Grozin:2014hna, Grozin:2015kna}.  Precisely, when expressed in terms of the light-like cusp replacing the coupling constant, the cusp anomalous dimension gives rise to an universal function $\Omega(K,\phi) $ that is independent of the number of fermion or scalar fields in the theory.

It is easy to prove that up to three loops the universal behaviour is also present  in $\mathcal{N}=2$ SCQCD.  With respect to the $\mathcal{N}=4$ SYM case, the cusp anomalous dimension gets the additional $\zeta(3)$ term in equation \eqref{Gfinal}  at three loops, which produces a corresponding term in the light--like cusp expansion  \eqref{lightlikecusp}.   Then one can invert  \eqref{lightlikecusp}  to express the coupling $g^2$ as a perturbative expansion in  $K$ and substitute this expansion back in the full cusp $\Gamma_{\text{cusp}}(g^2,N,\varphi)$ to obtain the function  $\Omega(K,\phi)$.  The additional $\zeta(3)$ terms coming from the genuine $\Gamma_{\text{cusp}}(g^2,N,\varphi)$ and from the substitution of the  expansion $g^2(K)$  trivially cancel, producing the same universal function  as derived in $\mathcal{N}=4$ SYM. In \cite{Grozin:2017css, Moch:2017uml} it was shown that at four loops the universality is in general violated. 

\section{Beyond three loops}
\label{section6}

The computational framework we have set up in this paper can be arguably extended to higher loops where some very non--trivial checks can be performed, especially on the existence of a universal behaviour shared by $\mathcal{N}=2$ SCQCD and  $\mathcal{N}=4$  SYM.

Let us here summarize the present understanding of this universal behaviour. 
First of all, it has been suggested in \cite{Gadde:2012rv} and then substantiated in \cite{Pomoni:2013poa} that the closed $SU(2,1|2)$ subsector of $\mathcal{N}=2$ SCQCD  inherits integrability from $\mathcal{N}=4$ SYM, since its Hamiltonian can be essentially  obtained from the $\mathcal{N}=4$  SYM one by substituting the coupling constant $g^2$ with an effective coupling $f(g^2)$. The explicit form of $f(g^2)$ was first derived in \cite{Mitev:2015oty} by comparing the exact results available from localization for circular BPS Wilson loops in  $\mathcal{N}=4$  SYM and $\mathcal{N}=2$ SCQCD. In the conventions of \cite{Mitev:2015oty} the first few orders in the weak coupling expansion read \footnote{In order to compare with our results we should substitute $g^2\to\frac{g^2N}{16\pi^2}$.}
\begin{align} \label{fexpansion}
f(g^2) & = g^2 -12 \zeta(3) g^6 + 120 \zeta(5) g^8 + \bigg(- 1120  \zeta(7) +  80 \zeta(2)\zeta(5) + 288 \zeta(3)^2 \bigg) g^{10} + \dots
\end{align}

It is  an interesting open problem to first understand the origin of this substitution rule and at the same time to test to what extent it is universal when applied to other observables that can be entirely built using fields from the $\mathcal{N}=2$ vector multiplet.  

In \cite{Mitev:2014yba} it was  conjectured that the effective coupling  $f(g^2)$ could be interpreted as the relative finite renormalization of the gluon  propagator of the two models, enforcing the argument presented in \cite{Pomoni:2013poa}. This proposal was supported by some diagrammatic checks of the coefficients of  \eqref{fexpansion}  up to order $g^8$ \cite{Mitev:2014yba}.   Nevertheless, first in \cite{Mitev:2014yba} and then in \cite{Mitev:2015oty},  it was noticed that this interpretation can be hardly extended at higher orders, because of the presence of terms that cannot be generated by purely massless  two-point integrals.  The $\zeta(2)\zeta(5) g^{10}$ contribution in  expansion \eqref{fexpansion} is the first example of such kind of terms, which ask for a clear interpretation.

Concerning the generalization of the substitution rule to other physical quantities, in \cite{Mitev:2015oty} a similar analysis was applied for extracting $f_B(g^2)$ from the comparison of the Bremsstrahlung functions of the two models, as computed from the Wilson loop expectation values on the ellipsoid.  In this case the effective coupling  $f_B(g^2)$ slightly differs from the one in  \eqref{fexpansion} starting at order $g^{10}$. In \cite{Mitev:2015oty} the discrepancy was explained as a consequence of scheme dependence in the choice of the relative infrared regulators. Once again, a term containing $\z(2)\z(5)$ appears, which cannot be explained if $f_B(g^2)$ has to be interpreted as the finite relative renormalization of gauge propagators, without resorting to coupling the model to curved space \cite{Mitev:2015oty}. Moreover, computing the Bremsstrahlung function from the two--point function of the stress energy tensor and the 1/2 BPS Wilson loop,  in \cite{Fiol:2015mrp} it was argued that in general for $\mathcal{N}=2$ theories with a single gauge group the substitution rule may be not working. 

The validity of the substitution rule is made even more obscured by the results on the direct computation  of purely adjoint scattering amplitudes in $\mathcal{N}=2$ SCQCD. In fact, it has been shown \cite{Leoni:2015zxa} that the amplitude/Wilson loop duality is broken already at two loops, displaying a qualitatively different functional dependence on the kinematic variables with respect to the $\mathcal{N}=4$ SYM amplitude. This arises even deeper questions about the integrability of the $SU(2,1|2)$  subsector, if the amplitude/Wilson loop duality has to be considered as direct consequence of integrability, like in $\mathcal{N}=4$ SYM \footnote{See the conclusions in \cite{Mitev:2015oty} for a discussion on possible ways out.}.

One way to shed some more light on the validity of the substitution rule and, in particular, on the actual origin of discrepancy terms of the form $\z(2)\z(5)$ would be a direct computation of the difference of the Bremsstrahlung  functions at higher orders, along the lines introduced in this paper.  Our  $SU(N)$ computation, preceded by the ones in \cite{Andree:2010na, Fiol:2015spa}, confirms the validity of the substitution rule \eqref{fexpansion} up to order $g^6$. From the diagrammatic analysis it is also clear how to associate the $\zeta(3) g^6$ term to diagrams containing propagator corrections.

At higher orders the situation is more intricate, but the use of the HQET techniques seems to be promising. At first,   the HQET integrals arise naturally as massive integrals, due to the presence of the heavy quark contour propagators. Indeed  using inversion transformations it is easy to map massive on shell propagator type integrals to HQET integrals, a procedure which has been used for QCD/HQET matching \cite{Broadhurst:1994se, Czarnecki:1997dz}. We consider for instance, as candidates for the production of $\z(2)\z(3)$ or  $\z(2)\z(5)$ terms, the massive propagator integrals introduced in  \cite{Mitev:2015oty}. Following for instance \cite{Grozin:2003ak}, inversion relations can be used to map such integrals to corresponding HQET versions

\begin{align}
\begin{minipage}{8cm}\vspace{-0.8cm}
\includegraphics[scale=0.3]{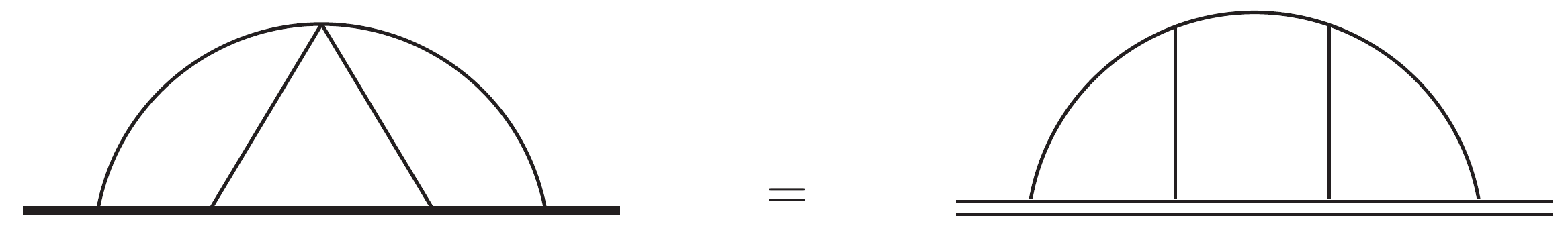}
\end{minipage} &\hspace{-0.4cm} \sim \,\,\,\,\, -5 \zeta (5) +12  \zeta (2)  \zeta (3)  \nonumber
\\[7mm]
\begin{minipage}{8cm}\vspace{-0.8cm}
\includegraphics[scale=0.3]{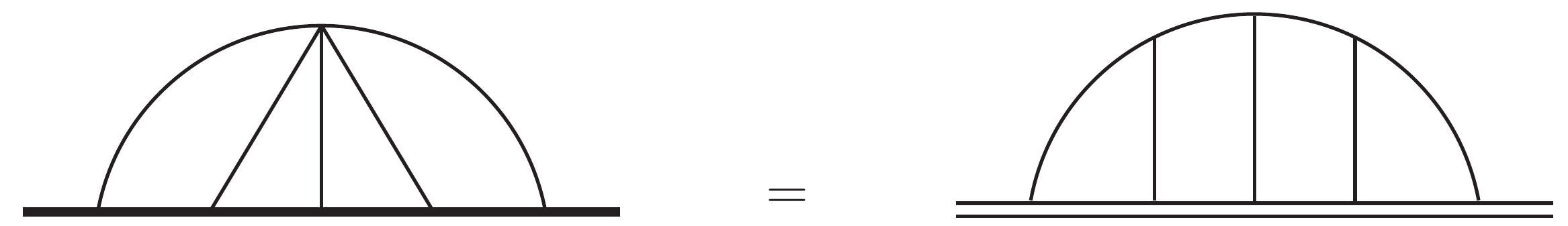}
\end{minipage} &  \hspace{-0.4cm} \sim \,\,\,\,\, -14 \zeta (7) -12  \zeta (3)  \zeta (4) +36  \zeta (2)  \zeta (5)  \nonumber
\end{align}

Here we indicate the massive propagators with a thick solid line and the WL contour with a double line.
In this way we are left with two examples of finite three and four loop HQET integrals containing $\z(2)\z(3)$ or  $\z(2)\z(5)$ terms.  Now the result of the integrals in our examples is finite, thus they cannot directly produce contributions to the cusp at three and four loops. Nevertheless,  it is easy to embed these HQET topologies in higher order diagrams,  producing poles potentially contributing to the cusp anomalous dimension.  For example we could proceed as follows

\begin{align}
\begin{minipage}{8.2cm}\vspace{-0.8cm}
\includegraphics[scale=0.23]{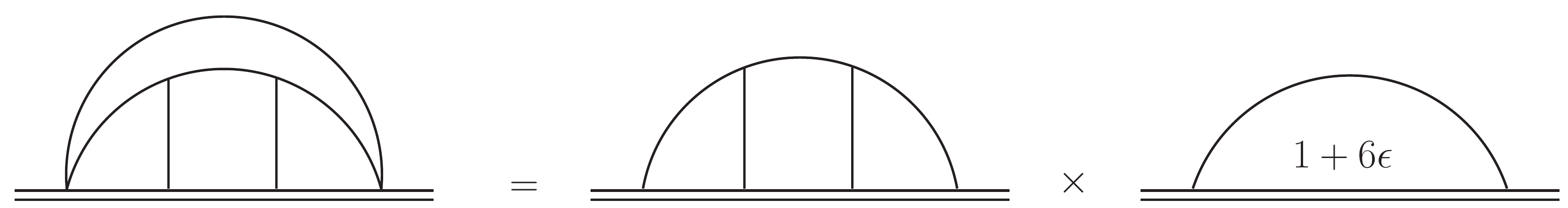}
\end{minipage} &\hspace{-0.3cm} \sim  \big(-5 \zeta (5) +12  \zeta (2)  \zeta (3) \big) \frac{1}{\e} \nonumber
\\[7mm]
\begin{minipage}{8.2cm}\vspace{-0.8cm}
\includegraphics[scale=0.23]{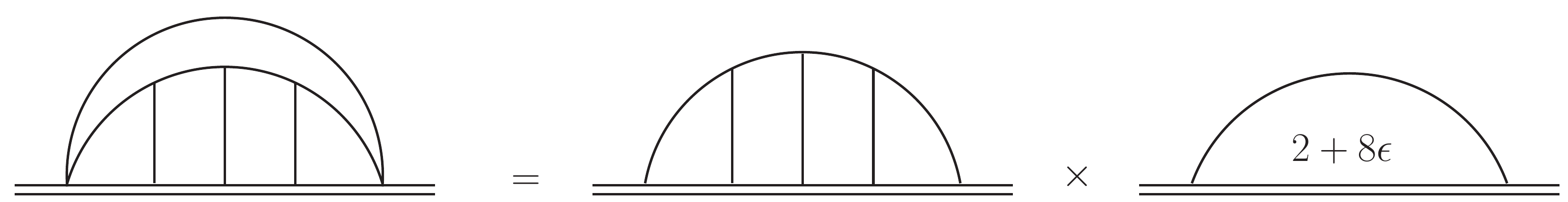}
\end{minipage} &  \hspace{-0.3cm} \sim \big(-14 \zeta (7) -12  \zeta (3)  \zeta (4) +36  \zeta (2)  \zeta (5) \big) \frac{1}{\e}  \nonumber
\end{align}
The integrals can be evaluated by factorization, reducing them to the product of our initial integrals and a one loop HQET bubble with a non-trivial index on the heavy line. 

Therefore, we conclude that in the HQET formalism  terms such as $\zeta (2)  \zeta (3)$ and  $\z(2)\z(5)$   can arise quite naturally in the expansion of the Bremsstrahlung function from the standard flat space computation of the cusped Wilson loop expectation value. In particular,  there is no need to introduce mass regulators, beside the usual IR cutoff $\delta$ that eventually drops out from the final result.
It is also natural to expect that some of these terms  survive once  taking the difference between  $\mathcal{N}=2$ SCQCD and $\mathcal{N}=4$ SYM, as predicted by the matrix model results. 

The interesting point is to understand whether these terms in the difference can be interpreted in terms of the substitution of an effective coupling given by  the finite different renormalization of the two point functions of the models, as advocated in \cite{Pomoni:2013poa, Mitev:2014yba}. Since we are working in component formalism, we expect that this claim should imply that at least some of these terms should originate from propagator type insertions into the cusp line.   At first glance, integrals such as the ones discussed above seem to originate from topologies which could hardly be associated to  propagator type diagrams.  However, without an explicit derivation we cannot draw any definite conclusion and therefore a direct calculation of the Bremsstrahlung function at four and five loops is  mandatory.
 
\vskip 50pt

\noindent
{\bf Acknowledgments}

\vskip 10pt
\noindent
We thank Luca Griguolo, Elli Pomoni and Domenico Seminara for useful discussions. 
This work has been supported in part by MIUR -- Italian Ministero dell'Istruzione, Universit\`a e Ricerca, and by INFN -- Istituto Nazionale di Fisica Nucleare through the ``Gauge Theories, Strings, Supergravity'' (GSS) research
project.

\newpage
\appendix

\section{Conventions }\label{appendix:conventions}

For $SU(N)$ gauge group we take the generators normalized as 
\begin{equation}
 \mathrm{Tr}(T^a T^b)=\frac12 \delta^{ab}
\end{equation}
whereas the structure constants can be read from
\begin{align} 
\label{coloridentities}
[T^a,T^b] & = i f^{abc}T^c \\
\{T^a,T^b\} & = \frac{1}{N}\delta^{ab} +  d^{abc} T^c  \label{identity}
\end{align}

For both $\mathcal{N}=2$ SCQCD and $\mathcal{N}=4$ SYM theories we derive the actions in components by projecting the euclidean $\mathcal{N}=1$ superfield action presented in \cite{Leoni:2014fja}. 
This is based on the conventions of  \cite{Gates:1983nr}, which we stick to.
 
For $\mathcal{N}=2$ $SU(N)$ SCQCD the superspace action reads
\begin{align}\label{superaction}
S \!   &=  \! \! \int \! \! \d^4x \d^4\theta \bigg[  \tr\big(e^{-g V} \bar{\Phi}e^{g V}\Phi\big) + \bar{Q}^{ I}e^{g V}Q_{ I}  + \tilde{Q}^{ I}e^{-g V}\bar{\tilde{Q}}_{ I} \bigg] \non \\ & + \frac{1}{g^2} \int\d^4x\d^2\theta \   \tr\big(W^{\alpha}W_{\alpha}\big)   \\
&+i g \int\d^4x\d^2\theta  \ \tilde{Q}^{ I} \Phi Q_{ I}
-i g \int\d^4x\d^2\bar\theta  \  \bar Q^{ I}\bar\Phi \bar{\tilde{Q}}_{ I}  \non
\end{align}
where $W_\alpha = i \bar{D}^2(e^{-g V}D_{\alpha}e^{g V})$ is the superfield strength of the $\mathcal{N}=1$ vector superfield $V$. The definition of the superspace covariant derivative $D_{\alpha}$ can be found in \cite{Gates:1983nr}. The $\mathcal{N}=1$ chiral superfield $\Phi$ transforms in the adjoint representation of $SU(N)$  and combines with $V$ into a $\mathcal{N}=2$ vector multiplet. The quark chiral scalar superfields $Q_{ I}$ and $\tilde{Q}^{ I}$ with  $I=1,\dots,N_f$ transform respectively in the fundamental and antifundamental representations of $SU(N)$ and together build up a  $\mathcal{N}=2$ hypermultiplet. At the critical value $N_f= 2N$ the action becomes exactly superconformal. 

We project action \eqref{superaction} down to components in the Wess-Zumino gauge and eliminate the auxiliary fields.  Defining the dynamical fields in terms of their $\mathcal{N}=1$ parents as  
\begin{align} \label{compdef}
& \Phi| = \sqrt{2} \, \phi \qquad D_{\alpha}\Phi| = \sqrt{2} \, \psi_{\alpha} \\
& Q_I| = q_I \qquad D_{\alpha}Q_I| = \lambda_{I \,\alpha} \\
& \frac{1}{2}[ \bar{D}_{\dot{\alpha}}, D_{\alpha}]V|  = \sqrt{2}  \, A_{\alpha\dot{\alpha}} =  (\sigma^\mu)_{\alpha \dot{\alpha}} A_\mu \qquad  i \bar{D}^2 D_{\alpha}V| = \sqrt{2} \, \eta_{\alpha} 
\end{align} 
where spinor and vector indices are converted using Pauli $\sigma$ matrices  
\begin{align}
& \sigma_{\m}^{\alpha \dot{\alpha}} \sigma^{\nu}_{\alpha \dot{\alpha}} = 2 \delta^{\nu}_{\mu} \qquad \sigma^{\m}_{\alpha \dot{\alpha}} \sigma_{\mu}^{\beta \dot{\beta}}= 2 \delta^{\beta}_{\alpha}\delta^{\dot{\beta}}_{\dot{\alpha}}
\end{align}
the final action in components reads
\begin{align}\label{N=2action}
S = & \int \!\! d^4x \,\bigg\{ 2 \, \textrm{Tr}\bigg[ i \psi^{\a}  (\sigma^{\m})_{\a}^{\,\,\,\dot{\b}} \mathcal{D}_{\m}\bar{\psi}_{\dot{\b}} + i \eta^{\a}  (\sigma^{\m})_{\a}^{\,\,\,\dot{\b}}  \mathcal{D}_{\m}\bar{\eta}_{\dot{\b}} \ \\
& - \frac{1}{4} F^{\mu \nu}F_{\mu \nu}  +\phi  \mathcal{D}^{\m}\mathcal{D}_{\m}\bar{\phi} -  \frac{g^2}{2} [\phi,\bar{\phi}][\phi,\bar{\phi}] + i g \sqrt{2}\, \bar{\psi}^{\dot{\a}}[ \bar{\eta}_{\dot{\a}}, \phi] - i g \sqrt{2}\, [ \bar{\phi}, \eta^{\a}] \psi_{\a}   \bigg]\non \\
&+ i \bar{\l}_{\dot{\b}}^{ I} (\sigma^{\m})_{\a}^{\,\,\,\dot{\b}} \mathcal{D}_{\m}\lambda^{\a}_{I}  +i \tilde{\lambda}^{\a I}  (\sigma^{\m})_{\a}^{\,\,\,\dot{\b}} \mathcal{D}_{\m}\bar{\tilde{\l}}_{\dot{\b} I} + \bar{q}^I \mathcal{D}^{\m} \mathcal{D}_{\m}q_I  + \tilde{q}^I  \mathcal{D}^{\m}\mathcal{D}_{\m} \bar{\tilde{q}}_I  \non \\
&  + ig \sqrt{2}\, ( \bar{\l}^{\dot{\a}I} \bar{\eta}_{\dot{\a}} q_I - \tilde{q}^I \bar{\eta}^{\dot{\a}} \bar{\tilde{\l}}_{\dot{\a} I}  ) - i g\sqrt{2}\,  ( \bar{q}^I \eta^{\a} \l_{\a I}  - \tilde{\l}^{\a I} \eta_{\a} \bar{\tilde{q}}_I)  + i g\sqrt{2}\, ( \tilde{\l}^{\a I} \psi_{\a} q_I - \bar{q}^I\bar{\psi}^{\dot{\a}}\bar{\tilde{\l}}_{\dot{\a}I} ) \non \\
& + i g\sqrt{2}\, ( \tilde{\l}^{\a I} \phi \l_{\a I}  -\bar{\l}^{\dot{\a}I}\bar{\phi}\bar{\tilde{\l}}_{\dot{\a}I} ) + i g\sqrt{2}\, (\tilde{q}^I \psi^{\a} \l_{\a I}  - \bar{\l}^{\dot{\a}I}\bar{\psi}_{\dot{\a}} \bar{\tilde{q}}_I) \non \\
& - g^2 \left[2\,\bar{q}^I \bar{\phi}\phi q_I +2\,\tilde{q}^I \phi \bar{\phi} \bar{\tilde{q}}_I + (\bar{q}^I q_J)( \tilde{q}^J \bar{\tilde{q}}_I  ) \right] 
  \non \\  
  &-  \frac{g^2}{4} \bigg[(\bar{q}^I q_J)(\bar{q}^J q_I) + ( \tilde{q}^I \bar{\tilde{q}}_J  )( \tilde{q}^J \bar{\tilde{q}}_I  ) - 2  (\bar{q}^I \bar{\tilde{q}}_J)( \tilde{q}^J q_I ) + 4\, \bar{q}^I [\phi,\bar{\phi}]q_I - 4\, \tilde{q}^I [\phi,\bar{\phi}]\bar{\tilde{q}}_I \bigg]  \bigg\} \nonumber 
\end{align}
The covariant derivatives are defined as
\begin{eqnarray}
\label{derivative}
 & \mathcal{D}^{\mu}q^{I}=\partial^{\mu}q^{I}- i g \, A^{\mu}q^{I} \\
 & \mathcal{D}^{\mu}\phi=\partial^{\mu}\phi- i g \, [A^{\mu},\phi] \nonumber
\end{eqnarray}

For $\mathcal{N}=4$ SYM theory, the $\mathcal{N}=1$ superspace description keeps manifest only a $SU(3)$ subgroup of the R-symmetry. The superspace action reads \cite{Gates:1983nr}
\begin{align}\label{superactionN=4}
S \!   &=  \! \! \int \! \! \d^4x \d^4\theta \, \tr\big(e^{-g V} \bar{\Phi}_I e^{g V}\Phi^I \big)  + \frac{1}{g^2} \int\d^4x\d^2\theta  \, \textrm{Tr}\big( W^{\alpha}W_{\alpha} \big)  \non \\
&+ \frac{i g}{3!} \int\d^4x\d^2\theta  \   \epsilon_{IJK}  \textrm{Tr} \big( \Phi^I [\Phi^J, \Phi^K]\big)+ h.c. 
\end{align}
where $I,J,K=1,2,3$ and all the fields transform in the adjoint representation of the gauge group $SU(N)$. Using field definitions similar to \eqref{compdef}, it is then straightforward to project down to component and obtain
\begin{align}\label{N=4action}
S = & \int \!\! d^4x \; 2\, \textrm{Tr}\bigg[  i \psi^{I \a}  (\sigma^{\m})_{\a}^{\,\,\,\dot{\b}} \mathcal{D}_{\m}\bar{\psi}_{I \dot{\b}} + i \eta^{\a}  (\sigma^{\m})_{\a}^{\,\,\,\dot{\b}}  \mathcal{D}_{\m}\bar{\eta}_{\dot{\b}}  - \frac{1}{4} F^{\mu \nu}F_{\mu \nu} +\phi^I  \mathcal{D}^{\m}\mathcal{D}_{\m}\bar{\phi_I}\non \\
&  + i g \sqrt{2}\, \bar{\psi}^{I \dot{\a}}[ \bar{\eta}_{\dot{\a}}, \phi_I] - i g\sqrt{2}\,[ \bar{\phi}^I, \eta^{\a}] \psi_{\a I}  +  i g \frac{\sqrt{2}}{2} \epsilon_{IJK}   [\phi^I, \psi^{\a J}] \psi^K_{\a } +i g \frac{\sqrt{2}}{2} \epsilon_{IJK}    [ \bar{\phi}^I, \bar{\psi}^{\dot{\a} J}] \bar{\psi}^K_{\dot{\a} } \non \\
& + g^2[\phi^I,\phi^J][\bar{\phi}^I,\bar{\phi}^J] - \frac{g^2}{2}[\phi^I,\bar{\phi}^I][\phi^J,\bar{\phi}^J]     \bigg]
\end{align}
From the previous actions the propagators in momentum space read
\begin{align}
\langle A_{\m}^a (x)A_{\n}^b (y) \rangle = & \; \delta^{ab} 
\int \frac{d^{4-2\e}p}{(2\pi)^{4-2\e}}e^{i p\cdot(x-y)}\frac{\delta_{\m\n}}{p^2}\\
\langle \bar{\phi}^a (x)\phi^b(y) \rangle = & \; \delta^{ab} 
\int \frac{d^{4-2\e}p}{(2\pi)^{4-2\e}}e^{i p\cdot(x-y)}\frac{1}{p^2}\\
\langle  \bar{q}^I(x)q_J(y) \rangle = & \langle  \bar{\tilde{q}}_J(x)\tilde{q}^I(y) \rangle = \delta^{I}_J \int \frac{d^{4-2\e}p}{(2\pi)^{4-2\e}}e^{i p\cdot(x-y)}\frac{1}{p^2}\\
\langle  \psi^{\a a }(x)\bar{\psi}_{\dot{\beta}}^b (y) \rangle = & \langle  \eta^{\a a }(x)\bar{\eta}_{\dot{\beta}}^b(y) \rangle  = \delta^{ab} \int \frac{d^{4-2\e}p}{(2\pi)^{4-2\e}}e^{i p\cdot(x-y)}\frac{(-p^{\m}) (\sigma_\m)^{\a}_{\,\,\,\dot{\beta}} }{p^2}\\
\langle  \l_J^{\a}(x)\bar{\l}^I_{\dot{\beta}}(y) \rangle = &\langle  \tilde{\l}^{\a I}(x)\bar{\tilde{\l}}_{\dot{\beta}J}(y) \rangle = \delta^{I}_J \int \frac{d^{4-2\e}p}{(2\pi)^{4-2\e}}e^{i p\cdot(x-y)}\frac{(-p^{\m}) (\sigma_\m)^{\a}_{\,\,\,\dot{\beta}} }{p^2}
\end{align}
The vertices entering the three--loop diagrams can be read directly from actions \eqref{N=2action} and \eqref{N=4action}.

\section{Results for the diagrams}\label{appendix:partial}

The three-loop Master Integrals introduced in section \ref{section4} are defined as follows

\begin{eqnarray}\label{G-int}
G_{a_1,...\,,a_{12}}=\int \frac{d^{d}k_{1/2/3}}{(2\pi)^{3d}} \frac{1}{P_1^{a_1}...\,P^{a_{12}}_{12}}
\end{eqnarray}
With
\begin{align}
P_1&= 1+2\widetilde{v}_1\cdot k_1 & P_7&=k_1^2 \nonumber\\
P_2&= 1+2\widetilde{v}_2\cdot k_1 & P_8&=k_2^2 \nonumber\\
P_3&= 1+2\widetilde{v}_1\cdot k_2 & P_9&=k_3^2\nonumber\\
P_4&= 1+2\widetilde{v}_2\cdot k_2 & P_{10}&=(k_1-k_2)^2 \nonumber\\
P_5&= 1+2\widetilde{v}_1\cdot k_3 & P_{11}&= (k_2-k_3)^2 \nonumber\\
P_6&= 1+2\widetilde{v}_2\cdot k_3 & P_{12}&= (k_1-k_3)^2 \nonumber
\end{align}

For specialized sets of $a_1, \dots , a_{12}$ indices these integrals can be computed analytically by {\em Mathematica} packages. Actually, for our porposes only the $\epsilon$ expansion of the result is necessary.  Omitting a common factor $\frac{e^{-3\epsilon\gamma_E}}{(4\pi)^{3d/2}}$ and stopping the expansion at the required order, the Master Integrals that enter our calculation read 
\begin{eqnarray}
& I_1\equiv G_{0,0,0,0,1,0,0,1,0,1,0,1}= \includegraphics[scale=0.3]{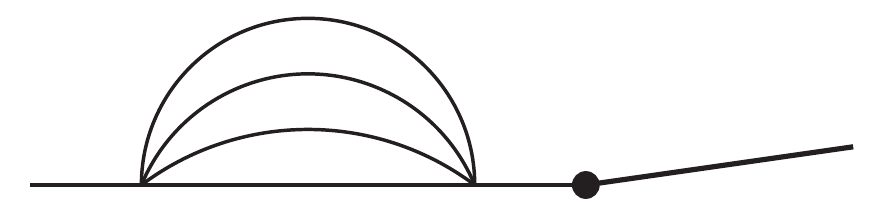} \nonumber\\
& = -\frac{1}{720 \epsilon}-\frac{137}{7200}-\frac{12019+325 \pi ^2 }{72000}\epsilon -\frac{874853+44525 \pi ^2-71000 \zeta[3]) }{720000}\epsilon^2 +\mathcal{O}(\epsilon^3)
\end{eqnarray}
\;
\begin{eqnarray}
& I_2 \equiv G_{0,0,0,0,1,1,0,1,0,1,0,1} = \includegraphics[scale=0.25]{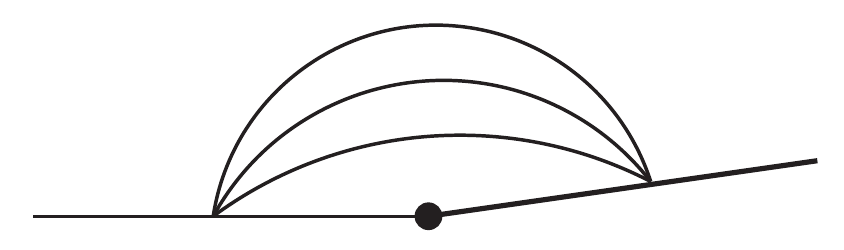} \nonumber\\
&=\frac{x \left(-1-8 x+8 x^3+x^4+12 x^2 \log(x)\right)}{144 (-1+x) (1+x)^5 \epsilon} \nonumber \\
& + \frac{x \left(-7+x \left(-59+3 \pi ^2 x+x^2 (59+7 x)\right)-9 x^2 \log(x) (-6+\log(x)-4 \log(1+x))+36 x^2 \text{Li}_2(-x)\right)}{72 (-1+x) (1+x)^5} \nonumber \\
& +\epsilon\left[\frac{x\left((-1+x) (1+x) (499+x (4400+499 x))+\pi ^2 (-13+x (-104+x (216+13 x (8+x))))\right)}{576 (-1+x) (1+x)^5} \right.\nonumber \\
& +\frac{ x^3 \left(\log(x) \left(207+7 \pi ^2-6 (9-\log(x)) \log(x)\right)+3 \left(\pi ^2+(6-\log(x)) \log(x)\right) \log(1+x)-6 (\log(-x)-\log(x)) \log(1+x)^2\right)}{48 (-1+x) (1+x)^5} \nonumber \\
& \left.+\frac{3 x^3 (3 \text{Li}_2(-x)-\text{Li}_3(-x)-2 \text{Li}_3(1+x)+\zeta(3))}{2 (-1+x) (1+x)^5}\right] \nonumber \\
& + \epsilon^2\left[\frac{x\left(1128 \pi ^4 x^2+10 \pi ^2 (-91+x (-767+x (621+13 x (59+7 x)))) \right) }{2880 (-1+x) (1+x)^5} \right. \nonumber \\
& + \frac{x^3\log(x) \left(648+42 \pi ^2-\log(x) \left(207+7 \pi ^2+3 (-12+\log(x)) \log(x)\right)\right) }{32 (-1+x) (1+x)^5} \nonumber \\
& + \frac{x^3 \left(54 \pi ^2+\log(x) \left(207+7 \pi ^2+6 (-9+\log(x)) \log(x)\right)\right) \log(1+x) }{8 (-1+x) (1+x)^5} \nonumber \\
& + \frac{x^3\left( 72 \left(\pi ^2-6 \log(-x)-(-6+\log(x)) \log(x)\right) \log(1+x)^2+96 (-\log(-x)+\log(x)) \log(1+x)^3\right) }{32(-1+x) (1+x)^5} \nonumber \\
& + \frac{x  \left(-3671-33586 x+33586 x^3+3671 x^4+284 \zeta(3)\right)}{576 (-1+x) (1+x)^5} \nonumber \\
& + \frac{ x^2 \left(18 (207+13 \pi ^2) x \text{Li}_2(-x)-648 x\left( (3+2 \log(1+x)) \text{Li}_3(-x)+6 \text{Li}_3(1+x)-\text{Li}_4(-x)+4 \text{Li}_4(1+x)+2 \text{S}_{2,2}(-x)\right) \right) }{144(-1+x) (1+x)^5} \nonumber \\
& + \left. \frac{x^2\left(568 \zeta(3)+x (1944-71 x (8+x)-204 \log(x)+1296 \log(1+x)) \zeta(3) \right) }{144 (-1+x) (1+x)^5} \right] + \mathcal{O}(\epsilon^3) \nonumber \\
\end{eqnarray}
\;
\begin{eqnarray}
&I_3\equiv G_{1,0,0,0,0,0,0,1,1,1,0,1} = \includegraphics[scale=0.3]{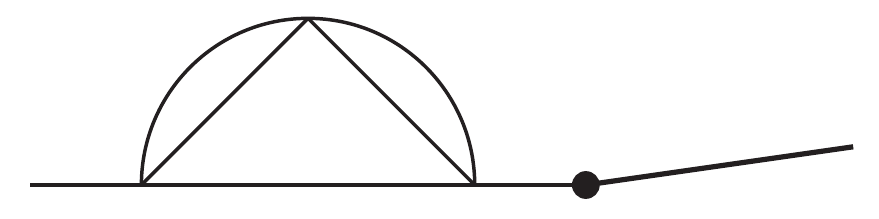} \nonumber\\
& = -\frac{1}{18 \epsilon^2}-\frac{2}{3 \epsilon}-\frac{16}{3}-\frac{13 \pi ^2}{72} - \frac{656+39 \pi ^2-65 \zeta[3]}{18} \epsilon + \mathcal{O}(\epsilon^2) \nonumber \\
\end{eqnarray}
\;
 \begin{eqnarray}
&  I_4 \equiv G_{1,1,0,0,0,0,0,1,1,1,0,1} = \includegraphics[scale=0.25]{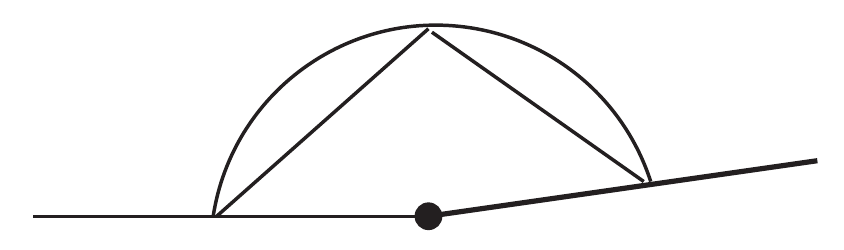}   \nonumber\\
& = \frac{x \left(-1+x^2+2 x \log(x)\right)}{3 (-1+x) (1+x)^3 \epsilon^2} + \frac{x \left(-13+x \left(\pi ^2+13 x\right)+x \log(x) (14-3 \log(x)+12 \log(1+x))+12 x \text{Li}_2(-x)\right)}{3 (-1+x) (1+x)^3 \epsilon}  \nonumber\\
&+\frac{x^2 \left( \log(x) \left(132+7 \pi ^2+6 (-7+\log(x)) \log(x)\right)+12 \left(3 \pi ^2+(14-3 \log(x)) \log(x)\right) \log(1+x)+72 (-\log(-x)+\log(x)) \log(1+x)^2\right) }{6 (-1+x) (1+x)^3}  \nonumber \\
&+ \frac{x \left(444 (-1+x^2)+\pi ^2 (-13+x (28+13 x)) \right) }{12 (-1+x) (1+x)^3} + \frac{4x^2 \left(7 \text{Li}_2(-x)-3 \text{Li}_3(-x)-6 \text{Li}_3(1+x)+3 \zeta(3) \right) }{ (-1+x) (1+x)^3}  \nonumber \\
&+ \epsilon \left[ \frac{x\left( 188 \pi ^4 x+15940 \left(-1+x^2\right)+\pi ^2 \left(-845+660 x+845 x^2\right)\right)}{60 (-1+x) (1+x)^3} \right.  \nonumber \\
&+ \frac{x^2 \log(x) \left(1048+98 \pi ^2-3 \log(x) \left(132+7 \pi ^2+\log(x) (-28+3 \log(x))\right)\right)}{12 (-1+x) (1+x)^3} \nonumber \\
& +\frac{x^2 \left(42 \pi ^2+\log(x) \left(132+7 \pi ^2+6 (-7+\log(x)) \log(x)\right)\right) \log(1+x)}{(-1+x) (1+x)^3}  \nonumber \\
&+ \frac{x^2\left( 72 \left(3 \pi ^2-14 \log(-x)+(14-3 \log(x)) \log(x)\right) \log(1+x)^2+288 (-\log(-x)+\log(x)) \log(1+x)^3\right)}{12 (-1+x) (1+x)^3} \nonumber \\
&+\frac{20x^2\left(\left(396+39 \pi ^2\right) \text{Li}_2(-x)-36 ((7+6 \log(1+x)) \text{Li}_3(-x)+14 \text{Li}_3(1+x)-3 \text{Li}_4(-x)+12 \text{Li}_4(1+x)+6 \text{S}_{2,2}(-x)) \right)}{60 (-1+x) (1+x)^3} \nonumber \\
&+\left.  \frac{x\left(1300 \zeta(3) + 20x (252-65 x-22 \log(x)+216 \log(1+x)) \zeta(3) \right)}{60 (-1+x) (1+x)^3} \right] + \mathcal{O}(\epsilon^2) \nonumber 
\end{eqnarray}

We can now write the contribution of every single diagram in figure \ref{diag} in terms of these Master Integrals.
Omitting a common factor $\frac{g^6(N^2-1)(N^2+1)}{2 N}$, from the insertion of corrected gauge propagators we have   

\begin{eqnarray}
 & (a)= \left[\frac{4 (-7+3 d) \left(9 d^5 (1+x (8+x (-2+x (8+x))))-64 (35+x (241+x (25+x (241+35 x)))) \right)}{3 (-5+d) (-4+d)^3 (-3+d) (-1+d) x^2} \right. \nonumber \\
 & + \frac{4 (-7+3 d) \left( -3 d^4 (47+x (388+x (-102+x (388+47 x))))+2 d^3 (425+x (3574+x (-830+x (3574+425 x))))\right)}{3 (-5+d) (-4+d)^3 (-3+d) (-1+d) x^2} \nonumber \\
 & + \left. \frac{4 (-7+3 d) \left( 8 d (469+x (3698+x (-190+x (3698+469 x))))-4 d^2 (631+x (5264+x (-850+x (5264+631 x))))\right)}{3 (-5+d) (-4+d)^3 (-3+d) (-1+d) x^2}\right]I_1 \nonumber \\
 & +\left[\frac{(80+9 (-6+d) d) (1+x)^4 \left( 224-308 d+160 d^2-37 d^3+3 d^4\right)}{3 (-5+d) (-4+d)^3 (-3+d) (-1+d) x^3}\right. \nonumber \\
 & +\left.\frac{(80+9 (-6+d) d) (1+x)^4 \left(6 (-4+d) (-3+d) (4+(-8+d) d) x+(-2+d) (-112+d (98+d (-31+3 d))) x^2 \right)}{3 (-5+d) (-4+d)^3 (-3+d) (-1+d) x^3} \right]I_2 \nonumber \\
 & + \left[\frac{12 (-3+d) \left(d \left(1+6 x+x^2\right)-4 \left(2+5 x+2 x^2\right)\right)}{(-5+d) (-4+d)^2 (-1+d) x}\right]I_3 \nonumber \\
 & + \left[\frac{(1+x)^2 \left(3 d^2 \left(1-6 x+x^2\right)+80 \left(1-3 x+x^2\right)-2 d \left(17-66 x+17 x^2\right)\right)}{(-5+d) (-4+d)^2 (-1+d) x^2}\right]I_4
\end{eqnarray}
\,
\begin{eqnarray}
 & (b) = \left[\frac{8 (-7+3 d) \left(\frac{1}{x}+x\right) (3 d (1+x (6+x))-2 (5+x (28+5 x)))}{(-4+d) (-3+d) x}\right]I_1 \nonumber \\
 & +\left[\frac{2 (80+9 (-6+d) d) (1+x)^4 \left(\frac{1}{x}+x\right)}{(-4+d) (-3+d) x^2}\right]I_2
\end{eqnarray}
\,
\begin{eqnarray}
& (c)= \left[-\frac{4 (-7 + 3 d)(9 d^3 (1+x (8+x (-2+x (8+x))))-16 (35+x (241+x (25+x (241+35 x)))))}{(-5 + d) (-4 + d)^2 (-3 + d) x^2}\right. \nonumber \\
& \left. +\frac{4 (-7 + 3 d)(3 d^2 (35+x (268+x (-30+x (268+35 x))))-d (418+2 x (1522+x (10+x (1522+209 x)))))}{(-5 + d) (-4 + d)^2 (-3 + d) x^2}\right]I_1 \nonumber \\
& +\left[-\frac{(80+9 (-6+d) d) (1+x)^4 \left(3 d^2 (1+x)^2+8 (7+x (9+7 x))-d (25+x (42+25 x))\right)}{(-5+d) (-4+d)^2 (-3+d) x^3}\right]I_2
\end{eqnarray}
\,
\begin{eqnarray}
& (d)=\left[-\frac{2 (-7 + 3 d)(-160 \left(1+x^2\right) (5+x (28+5 x))+9 d^3 (1+x (8+x (-2+x (8+x)))))}{3 (-5+d) (-4+d)^2 (-3+d) x^2}\right. \nonumber \\
& \left.-\frac{2 (-7 + 3 d)(-6 d^2 (19+x (134+x (6+x (134+19 x))))+8 d (65+x (406+x (100+x (406+65 x)))))}{3 (-5+d) (-4+d)^2 (-3+d) x^2} \right]I_1 \nonumber \\
& +\left[-\frac{(-10+3 d) (-8+3 d) (1+x)^4 \left(3 d^2 (1+x)^2+80 \left(1+x^2\right)-4 d (7+x (6+7 x))\right)}{6 (-5+d) (-4+d)^2 (-3+d) x^3}\right]I_2
\end{eqnarray}
\,
\begin{eqnarray}
& (e)=  \left[\frac{8 (-7+3 d) \left(9 d^5 (1+x (2+x (10+x (2+x))))-3 d^4 (53+x (148+x (390+x (148+53 x))))\right)}{3 (-5+d) (-4+d)^3 (-3+d) (-1+d) x^2}\right. \nonumber \\
& + \left. \frac{8 (-7+3 d) \left(-32 (85+x (296+x (350+x (296+85 x))))+16 d (296+x (1045+x (1315+x (1045+296 x))))\right)}{3 (-5+d) (-4+d)^3 (-3+d) (-1+d) x^2}\right. \nonumber \\
& \left.  \frac{8 (-7+3 d) \left( 2 d^3 (518+x (1663+x (3046+x (1663+518 x))))-4 d^2 (799+x (2747+x (3986+x (2747+799 x))))\right)}{3 (-5+d) (-4+d)^3 (-3+d) (-1+d) x^2}\right]I_1 \nonumber \\
& +\left[\frac{2 (80+9 (-6+d) d) (1+x)^4 \left( 272-392 d+202 d^2-43 d^3+3 d^4\right)}{3 (-5+d) (-4+d)^3 (-3+d) (-1+d) x^3}\right.\nonumber \\
& +\left.\frac{2 (80+9 (-6+d) d) (1+x)^4 \left( -6 (-4+d) (-3+d) (8+d (-7+2 d)) x+(-2+d) (-136+d (128+d (-37+3 d))) x^2\right)}{3 (-5+d) (-4+d)^3 (-3+d) (-1+d) x^3}\right]I_2\nonumber \\
& +\left[\frac{6 (-3+d) (8+(-5+d) d) (-4 (2+x) (1+2 x)+d (1+x (6+x)))}{(-5+d) (-4+d)^2 (-1+d) x}\right]I_3 \nonumber \\
& + \left[\frac{(-10+3 d) (8+(-5+d) d) (1+x)^2 \left(-8+d-6 (-4+d) x+(-8+d) x^2\right)}{2 (-5+d) (-4+d)^2 (-1+d) x^2}\right]I_4
\end{eqnarray}
\,
\begin{eqnarray}
 & (f)= \left[\frac{4 (-7+3 d) \left(9 d^3 (1+x)^4-3 d^2 (33+x (132+x (166+33 x (4+x))))\right)}{3 (-5+d) (-4+d) (-3+d) (-1+d) x^2}\right. \nonumber \\
 & \left.+\frac{4 (-7+3 d) \left(4 d (83+x (316+x (370+x (316+83 x))))-4 (85+x (296+x (350+x (296+85 x))))\right)}{3 (-5+d) (-4+d) (-3+d) (-1+d) x^2} \right]I_1 \nonumber\\
 & +\left[\frac{(80+9 (-6+d) d) (1+x)^4 \left(34-23 d+3 d^2-6 (-4+d) (-3+d) x+(-2+d) (-17+3 d) x^2\right)}{3 (-5+d) (-4+d) (-3+d) (-1+d) x^3}\right]I_2 \nonumber \\ 
& +\left[\frac{6 (-3+d)^2 (-4 (2+x) (1+2 x)+d (1+x (6+x)))}{(-5+d) (-4+d) (-1+d) x}\right]I_3 \nonumber \\
& +\left[\frac{(-3+d) (-10+3 d) (1+x)^2 \left(-8+d-6 (-4+d) x+(-8+d) x^2\right)}{2 (-5+d) (-4+d) (-1+d) x^2}\right]I_4
 \end{eqnarray}
\,
\begin{eqnarray}
& (g)= \left[\frac{8 (-7+3 d)\left(9 d^3 (1+x (8+x (-2+x (8+x))))+8 (5+x (-62+x (100+x (-62+5 x)))) \right) }{3 (-5+d) (-4+d) (-3+d) (-1+d) x^2}\right. \nonumber \\
& \left. + \frac{8 (-7+3 d)\left(-3 d^2 (23+x (196+x (-54+x (196+23 x))))+2 d (59+x (646+x (-290+x (646+59 x)))) \right) }{3 (-5+d) (-4+d) (-3+d) (-1+d) x^2}\right]I_1 \nonumber \\
& +\left[\frac{2 (80+9 (-6+d) d) (1+x)^4 \left(3 d^2 (1+x)^2-4 (1+(-18+x) x)-d (13+x (42+13 x))\right)}{3 (-5+d) (-4+d) (-3+d) (-1+d) x^3}\right]I_2 \nonumber \\
& +\left[\frac{24 (-3+d) (-4 (2+x) (1+2 x)+d (1+x (6+x)))}{(-5+d) (-4+d) (-1+d) x}\right]I_3 \nonumber \\
& +\left[\frac{2 (-10+3 d) (1+x)^2 \left(-8+d-6 (-4+d) x+(-8+d) x^2\right)}{(-5+d) (-4+d) (-1+d) x^2}\right]I_4
 \end{eqnarray}
\,
\begin{eqnarray}
 & (h) =  \left[-\frac{64 (-7+3 d) \left(9 d^3 (-1+x)^2 x-2 d (1+(-12+x) x) (8+x (-17+8 x))\right)}{3 (-5+d) (-4+d)^2 (-3+d) x^2}\right. \nonumber \\
 & \left.-\frac{64 (-7+3 d) \left(3 d^2 (1+x (-25+x (52+(-25+x) x)))+4 (5+x (-62+x (100+x (-62+5 x))))\right)}{3 (-5+d) (-4+d)^2 (-3+d) x^2} \right]I_1 \nonumber \\
 & +\left[-\frac{16 (80+9 (-6+d) d) (1+x)^4 \left(-2+d+3 (-4+d) (-3+d) x+(-2+d) x^2\right)}{3 (-5+d) (-4+d)^2 (-3+d) x^3}\right]I_2
\end{eqnarray}
\,
Similarly, from the insertion of corrected  scalar propagators, omitting a common factor $\frac{g^6(N^2-1)(N^2+1)}{2 N} \cos\theta$ we have
\begin{eqnarray}
 &(i) = \left[-\frac{32 (-7+3 d) (16+d (-11+2 d)) (3 d (1+x (6+x))-2 (5+x (28+5 x)))}{(-4+d)^3 (-3+d) x}\right]I_1\nonumber \\
& -\left[\frac{8 (-10+3 d) (-8+3 d) (16+d (-11+2 d)) (1+x)^4}{(-4+d)^3 (-3+d) x^2}\right]I_2 \nonumber \\
& -\left[\frac{96 (-3+d)^2}{(-4+d)^2}\right]I_3 -\left[\frac{8 (-3+d) (-10+3 d) (1+x)^2}{(-4+d)^2 x}\right]I_4
 \end{eqnarray}
\,
\begin{eqnarray}
& (j)= \left[-\frac{16 (-7+3 d) (3 d (1+x (6+x))-2 (5+x (28+5 x)))}{(-4+d) (-3+d) x}\right]I_1 \nonumber \\
& - \left[\frac{4 (80+9 (-6+d) d) (1+x)^4 }{(-4+d) (-3+d) x^2}\right]I_2
\end{eqnarray}
\,
\begin{eqnarray}
 & (k) = \left[-\frac{64 (-2+d) (-7+3 d) (3 d (1+x (6+x))-2 (5+x (28+5 x))) }{(-4+d)^2 (-3+d) x}\right]I_1\nonumber \\
 & - \left[\frac{16 (-2+d) (80+9 (-6+d) d) (1+x)^4 }{(-4+d)^2 (-3+d) x^2}\right]I_2
\end{eqnarray}

\bibliographystyle{JHEP}

\bibliography{Biblio}

\end{document}